\documentclass[reprint, twocolumn, superscriptaddress, pra]{revtex4-2}
\usepackage{bm, amsmath, amsfonts, amssymb}
\usepackage[italicdiff]{physics}
\usepackage{braket}

\usepackage{subfigure}
\usepackage{color}
\usepackage{graphicx}

\usepackage{qcircuit}
\usepackage{url}
\usepackage{bm}
\usepackage[breaklinks=true]{hyperref}

\newcommand{\mr}[1]{\mathrm{#1}}

\begin{document}
\title{Non-normal Hamiltonian dynamics in quantum systems and its realization on quantum computers}

\author{Nobuyuki Okuma}
\email{okuma@hosi.phys.s.u-tokyo.ac.jp}
\affiliation{
 Yukawa Institute for Theoretical Physics, Kyoto University, Kyoto 606-8502, Japan
}

\author{Yuya O. Nakagawa}
\affiliation{QunaSys Inc., Aqua Hakusan Building 9F, 1-13-7 Hakusan, Bunkyo, Tokyo 113-0001, Japan}

\date{\today}
\begin{abstract}
The eigenspectrum of a non-normal matrix, which does not commute with its Hermitian conjugate, is a central issue of non-Hermitian physics that has been extensively studied in the past few years.
There is, however, another characteristic of a non-normal matrix that has often been  overlooked: the pseudospectrum, or the set of spectra under small perturbations.
In this paper, we study the dynamics driven by the non-normal matrix (Hamiltonian) realized as a continuous quantum trajectory of the Lindblad master equation in open quantum systems and point out that the dynamics can reveal the nature of unconventional pseudospectrum of the non-normal Hamiltonian.
In particular, we focus on the transient dynamics of the norm of an unnormalized quantum state evolved with the non-normal Hamiltonian, which is related to the probability for observing the trajectory with no quantum jump.
We formulate the transient suppression of the decay rate of the norm due to the pseudospectral behavior and derive a non-Hermitian/non-normal analog of the time-energy uncertainty relation.
We also consider two methods to experimentally realize the non-normal dynamics and observe our theoretical findings on quantum computers: one uses a technique to realize non-unitary operations on quantum circuits and the other leverages a quantum-classical hybrid algorithm called variational quantum simulation.
Our demonstrations using cloud-based quantum computers provided by IBM Quantum exhibit the frozen dynamics of the norm in transient time, which can be regarded as a non-normal analog of the quantum Zeno effect.
\end{abstract}

\maketitle
\section{Introduction}
Recently, the spectral theory of non-Hermitian Hamiltonians has been extensively studied in various fields of physics \cite{Bender-98, Bender-02, Bender-review, Konotop-review, Christodoulides-review, Gong-18, KSUS-19,ZL-19,ashida-gong-20,Poli-15,Zeuner-15,Zhen-15,Zhou-18,Weimann-17,Xiao-17,St-Jean-17,Bahari-17,Harari-18,Bandres-18,Zhao-19,Hatano-Nelson-96,Hatano-Nelson-97,Lee-16,MartinezAlvarez-18,Torres-2019,YW-18-SSH,YSW-18-Chern,Kunst-18,YM-19,KOS-20,YZFH-19,Lee-19,OKSS-20, Zhang-19,Okuma-19,Borgnia-19,Brandenbourger-19-skin-exp,Ghatak-19-skin-exp,Helbig-19-skin-exp, Hofmann-19-skin-exp,Xiao-19-skin-exp,Weidemann-20-skin-exp,Brandenbourger-19-skin-exp,Ghatak-19-skin-exp,Xiao-19-skin-exp,Weidemann-20-skin-exp,Kozii-17,Yoshida-18,Yoshida-19,Bergholtz-19,Kimura-19,Okugawa-19,Budich-19,KBS-19,Bardarson-19,Ryu-20,Ye-20}.
One of the most interesting phenomena in non-Hermitian physics is the non-Hermitian skin effect \cite{Hatano-Nelson-96,Hatano-Nelson-97,Lee-16,MartinezAlvarez-18,Torres-2019,YW-18-SSH,YSW-18-Chern,Kunst-18,YM-19,KOS-20,YZFH-19,Lee-19,OKSS-20, Zhang-19,Okuma-19,Borgnia-19,Brandenbourger-19-skin-exp,Ghatak-19-skin-exp,Helbig-19-skin-exp, Hofmann-19-skin-exp,Xiao-19-skin-exp,Weidemann-20-skin-exp}, in which the bulk spectrum under the open boundary condition (OBC) is far from that under the periodic boundary condition (PBC) owing to the localized boundary modes.
Such non-Hermitian boundary modes, called the non-Bloch wavefunctions \cite{YW-18-SSH,YSW-18-Chern,Kunst-18,YM-19}, 
have shown to be related to non-Hermitian topology \cite{OKSS-20,Zhang-19}, and the symmetry-protected variants have also been proposed \cite{OKSS-20}.
Another important aspect of non-Hermitian physics is the non-diagonalizable nature at the special point of a parameter-dependent Hamiltonian, called the exceptional point \cite{Kozii-17,Yoshida-18,Yoshida-19,Bergholtz-19,Kimura-19,Okugawa-19,Budich-19,KBS-19}.

Strictly speaking, the unconventional behaviors in both of the above examples come from the non-normality of the Hamiltonian $H$ (i.e., $[H,H^\dagger]\neq0$).
For a non-normal matrix, not only the spectrum but also pseudospectrum, which is defined as the set of the spectra in the presence of small perturbations to the matrix, becomes nontrivial \cite{Trefethen}.
In contrast to normal matrices, including Hermitian ones, even a small perturbation can cause a drastic change of the spectrum, which means that the pseudospectrum is much larger than the neighborhood of the original spectrum.
While the notion of the pseudospectrum has often been overlooked, recent studies have shown its importance in various fields, including fluid mechanics \cite{Trefethen}, network science \cite{Asllani-2018}, and non-Hermitian topological phenomena \cite{Okuma-Sato-20,Okuma-Sato-21}.
In particular, the transient dynamics governed by a linear equation of a matrix is known to be described by its pseudospectrum~\cite{Trefethen}.
In other words, the presence of the non-normality should affect the amplification or the relaxation process towards the steady-state of the system.

In this paper, we study the dynamics of quantum systems driven by non-normal Hamiltonian matrices.
Our contributions are composed of theoretical and experimental parts.
In the theoretical part, we consider the non-normal Hamiltonian dynamics defined as the quantum trajectory with no quantum jump process under the Lindblad master equation~\cite{Lindblad}.
We focus on the fact that the probability for observing the trajectory with no quantum jump is related to the norm of quantum states evolved with the non-normal Hamiltonian. 
We formulate the dynamics of the probability (and the norm) in terms of the spectral theory of non-normal matrices.
Especially, we point out the possible transient suppression of the decay rate of the probability due to the pseudospectral behavior of the non-normal matrix.
Moreover, we derive a non-Hermitian/non-normal analog of the time-energy uncertainty relation, which gives a physical interpretation of the mathematical fact known in the spectral theory of non-normal matrices. 
In the experimental part, we think of two ways to realize the non-normal Hamiltonian dynamics on quantum computers~\cite{Nielsen2011} and perform actual experiments of them.
The first way is a direct implementation of the Lindblad equation using a technique to execute non-unitary operations on quantum circuits~\cite{Terashima-Ueda-03}.
The other one takes advantage of a quantum-classical algorithm called variational quantum simulation (VQS)~ \cite{Li2017efficient, Endo2020general}, which allows us to simulate the non-normal Hamiltonian dynamics within a pre-determined Hilbert space of trial wavefunctions.
The experiments for both methods performed on cloud-based quantum computers provided by IBM Quantum~\cite{IBMQ2021} yield the frozen dynamics of the norm and the wavefunction in transient time, which can be regarded as a non-normal analog of the quantum Zeno effect.

This paper is organized as follows.
In Sect.~\ref{sect: basics-of-non-normal}, we review the basic properties of the spectral theory of non-normal matrices.
Section~\ref{sect: non-normal quantum dynamics} is dedicated to the theoretical part of our paper.
In Sect.~\ref{sect: non-normal quantum dynamics}, we apply the spectral theory of non-normal matrices to non-Hermitian Hamiltonian dynamics that is defined as the continuous quantum trajectory of the Lindblad equation.
We relate the norm dynamics to the probability for the quantum trajectory with no quantum jump process.
We also propose the non-normal analog of the time-energy uncertainty relation in terms of the pseudospectral dynamics.
Sections~\ref{sect: theory of quantum circuits} and \ref{sect: experiment} consist of the experimental part of our paper.
In Sect.~\ref{sect: theory of quantum circuits}, we describe theories for the implementation of non-normal Hamiltonian dynamics on quantum computers.
In Sect. \ref{sect: experiment}, we present experimental results of the methods explained in Sect.~\ref{sect: theory of quantum circuits} on actual quantum computers.
We conclude our study in Sect.~\ref{sect: conclusion}.

\section{Basic properties of non-normal matrices\label{sect: basics-of-non-normal}}
In this section, we review the definitions and properties of several concepts in non-normal matrix theory.
In particular, we focus on the non-Hermitian skin effect as a typical example of a non-normal phenomenon, and relate it to important concepts of non-normal matrices including pseudospectrum.

\subsection{Spectra of non-normal matrices}
A matrix $H$ is normal if it is unitarily diagonalizable \cite{Trefethen}:
\begin{align}
    H=UDU^{-1},
\end{align}
where $D$ is a diagonal matrix of eigenvalues, and $U$ is a unitary matrix of a complete set of orthogonal eigenvectors.
An equivalent condition for normality is given by
\begin{align}
    [H,H^\dagger]=0.
\end{align}
Typical examples of normal matrices are Hermitian and unitary matrices, both of which are main targets of the conventional quantum mechanics.

A matrix $H$ is non-normal if it is not normal.
If it is diagonalizable, it is diagonalized only by a non-unitary matrix $P$ (i.e., $H=PDP^{-1}$).
Remarkably, the right and left eigenvectors of non-normal matrices are not always equivalent each other:
\begin{align}
    H|i\rangle&=E_i|i\rangle,\notag\\
    \langle\!\langle i|H&=E_i\langle\!\langle i|.\label{leftright}
\end{align}
Note that non-normal matrices are not always diagonalizable, as in the case of the following example:
\begin{align}
    H=
    \begin{pmatrix}
    0&0\\
    1&0
    \end{pmatrix}.
\end{align}

To see the difference between normal and non-normal matrices, let us consider the following $N\times N$ matrix that represents the Hatano-Nelson model without disorder \cite{Hatano-Nelson-96,Hatano-Nelson-97}:
\begin{equation}
H:=
\begin{pmatrix}
0&t-g&0&\cdots\\
t+g&0&t-g&\cdots\\
0&t+g&0&\cdots\\
\vdots&\vdots&\vdots&\ddots
\end{pmatrix}.
\label{Hatano-Nelson}
\end{equation}
In the language of the non-Hermitian tight-binding model,
$t\in\mathbb{R}$ and $g\in \mathbb{R}$ represent symmetric and asymmetric hopping terms on a one-dimensional lattice, respectively.
If we impose the PBC ($H_{(1,N)/(N,1)}=t\pm g$), the eigenspectrum is simply calculated by using the Fourier transform:
\begin{align}
    E_k=(t+g)e^{ik}+(t-g)e^{-ik},\label{PBCdispersion}
\end{align}
where $k$ is the crystal momentum, and the eigenvectors are given by delocalized plane waves.
Since the Fourier transform is a unitary transformation, $H$ under the PBC is a normal matrix.
If we impose the OBC ($H_{(1,N)/(N,1)}=0$), $H$ can be mapped to a Hermitian matrix $H'$ by using the imaginary gauge transformation $V_r$ \cite{Hatano-Nelson-96,Hatano-Nelson-97}:
\begin{align}
H' &:= V_{r}^{-1} H_{\rm OBC} V_{r}\notag\\
&=
\begin{pmatrix}
0&\sqrt{t^2-g^2}&\cdots\\
\sqrt{t^2-g^2}&0&\cdots\\
\vdots&\vdots&\ddots
\end{pmatrix},\label{similarity}
\end{align}
where $[V_r]_{i,j}=r^i\delta_{ij}$ with $r=\sqrt{(t+g)/(t-g)}$.
For simplicity, we have assumed $0<g<t$ in Eq. (\ref{similarity}).
Since the imaginary gauge transformation is a similarity transformation, which does not change the eigenspectrum of a finite matrix,
the eigenvalues of $H$ under the OBC are real, which are very different from the PBC ones (\ref{PBCdispersion}) on an ellipse in the complex plane.
Recently, this extreme sensitivity of the eigenspectrum against the boundary condition, which is called the non-Hermitian skin effect, has been extensively studied for various lattice models \cite{Hatano-Nelson-96,Hatano-Nelson-97,Lee-16,MartinezAlvarez-18,Torres-2019,YW-18-SSH,YSW-18-Chern,Kunst-18,YM-19,KOS-20,YZFH-19,Lee-19,OKSS-20, Zhang-19,Okuma-19,Borgnia-19,Brandenbourger-19-skin-exp,Ghatak-19-skin-exp,Helbig-19-skin-exp, Hofmann-19-skin-exp,Xiao-19-skin-exp,Weidemann-20-skin-exp}.
Remarkably, the eigenvectors, or $skin$ $modes$, are exponentially localized at the boundaries. In the case of the Hatano-Nelson model,
all the eigenstates are shown to be localized at one boundary by acting $V^{r}$ on the delocalized eigenstates of the Hermitian matrix $H'$.
Owing to the extreme non-normality, the left and right eigenvectors, defined in Eq. (\ref{leftright}), are localized at the opposite sides.
In this sense, these boundary modes are essentially different from the Hermitian boundary modes such as in topological insulators/superconductors \cite{Kane-review,Zhang-review}, while one of the authors has shown that the non-Hermitian skin modes originate from the non-Hermitian topology, and generalized the non-Hermitian skin effect to various symmetry classes and spatial dimensions \cite{OKSS-20}.

\subsection{Pseudospectra of non-normal matrices}
In the presence of the non-normality, the notion of the pseudospectrum becomes also important as well as the (eigen)spectrum. For $\epsilon>0$, the $\epsilon$-pseudospectrum $\sigma_\epsilon(H)$ is defined for a matrix $H\in\mathbb{C}^{N\times N}$ in the following three equivalent ways  \cite{Trefethen}:
\begin{itemize}
\item The set of $z\in\mathbb{C}$ such that $\|(z-H)^{-1}\|>\epsilon^{-1}$.
\item The set of $z\in\mathbb{C}$ such that $z\in\sigma(H+\eta)$ for some $\eta\in\mathbb{C}^{N\times N}$ with $\|\eta\|<\epsilon$.
\item The set of $z\in\mathbb{C}$ such that $\|(z-H)|v\rangle\|<\epsilon$ for some unit vector $|v\rangle\in\mathbb{C}^{N}$.
\end{itemize}
Here $\sigma(\cdot)$ is the spectrum of a matrix, and the norm $\|\cdot\|$ is defined as the conventional $2$-norm for a vector and the spectral norm for a matrix (i.e., the largest singular value), respectively. By construction, $\epsilon\rightarrow0$ corresponds to the conventional spectrum.
From the second definition, one can find that the pseudospectrum is a union of the spectra of the perturbed matrices.
Note that a complex number $z$ and the corresponding vector $|v\rangle$ in the third definition are almost an eigenvalue/eigenvector except for a deviation $\epsilon$. In this paper, we use the terminologies $pseudoeigenvalue$/$pseudoeigenvector$ for them. 

In the absence of non-normality, the $\epsilon$-pseudospectrum is simply given by the $\epsilon$-neighborhood of the unperturbed spectrum  \cite{Trefethen}:
\begin{align}
    \sigma_\epsilon(H)=\sigma(H)+\Delta_\epsilon:=\{z~|~{\rm dist}(z,\sigma(H))<\epsilon\},
\end{align}
where dist($\cdot,\cdot$) denotes the distance between two points in the complex plane.
In other words, a small perturbation to the matrix has a small effect on the spectral behavior of that matrix. In Hermitian/unitary physics, this fact is often used in the perturbation theory.

In the presence of the non-normality, on the other hand, the $\epsilon$-pseudospectrum can be greater than the $\epsilon$-neighborhood of the unperturbed spectrum  \cite{Trefethen}:
\begin{align}
     \sigma_\epsilon(H)\supset\sigma(H)+\Delta_\epsilon.
\end{align}
This relation implies that 
a small perturbation can drastically change the spectrum.
An extreme example is the pseudospectrum of the OBC Hatano-Nelson model~(\ref{Hatano-Nelson}), $H_\mr{OBC}$.
While the $\epsilon$-pseudospectrum of the mapped Hermitian Hamiltonian $H'$ in Eq.~\eqref{similarity} is given by the $\epsilon$-neighborhood of its own spectrum, that of the Hatano-Nelson model is larger than the $\epsilon$-neighborhood [Fig. \ref{fig: schematic pictures}(a)].
In the infinite-volume limit, it is known that the following relation holds \cite{Trefethen}:
\begin{align}
 \lim_{\epsilon\rightarrow0}\lim_{N\rightarrow\infty}\sigma_\epsilon(H_{\mr{OBC}})=\sigma(H_{\rm semi}),\label{semi-infinite}
\end{align}
where $H_{\rm semi}$ represents the infinite matrix (operator) that corresponds to the semi-infinite boundary condition, in which the boundary is at the left end of the semi-infinite lattice.
According to the Toeplitz index theorem \cite{Trefethen,Bottcher}, the semi-infinite spectrum in the right hand side of Eq.~\eqref{semi-infinite} is given by the PBC spectral curve [Eq.~\eqref{PBCdispersion}] in the infinite-volume limit together with all the points enclosed by the PBC curve with a nonzero winding number.
That is, 
the pseudospectrum of $H_\mr{OBC}$ for an infinitesimally small $\epsilon$ under the infinite-volume limit (with $0<g<t$) is a filled ellipse in the complex plane.
On the other hand, the spectrum of $H_\mr{OBC}$ becomes a line on the real axis under the infinite-volume limit. Therefore, the pseudospectrum of $H_\mr{OBC}$ is much larger than the $\epsilon$-neighborhood of the original spectrum.
Note that $\lim_{\epsilon\rightarrow0}$ and $\lim_{N\rightarrow\infty}$ does not commute in Eq. (\ref{semi-infinite}), and the opposite order gives the OBC exact spectrum.
The above discussion also holds for general Toeplitz matrices with short-range hoppings and other non-Hermitian tight-binding models in which the non-Hermitian skin effects occur \cite{Okuma-Sato-20}.

For a large but finite-size system, there is a threshold of $\epsilon$ for which the area of pseudospectrum drastically varies, and it is exponentially small with respect to the system size.
While Ref. \cite{OKSS-20} has shown the correspondence between the exact boundary zero modes of some Hermitian topological insulators/superconductors and skin modes of the $symmetry$-$protected$ non-Hermitian skin effects, Ref.~\cite{Okuma-Sato-20} has shown the correspondence between the quasi-zero modes with energy $\sim\epsilon$ of the former and the $\epsilon$-pseudospectrum of the latter. In this context, the above-mentioned exponentially small threshold can be related to the exponentially small quasi-zero energy due to the overlap of the boundary modes localized on the other sides in class-AIII topological insulators. 
By considering the correspondence in high dimensions, one can also consider the threshold proportional to the inverse of the system size \cite{Okuma-Sato-20}. 
Note that the exponentially-small threshold can also be understood in terms of the Lieb-Robinson bound (see Ref. \cite{Gong-18} for details).

\subsection{Long-time, short-time, and transient dynamics driven by non-normal Hamiltonian}
It is known that the dynamics driven by non-normal matrices is very different from that by normal matrices \cite{Trefethen}. Here we consider the Schr\"{o}dinger-type equation with $Hamiltonian$ matrix $H$ that allows non-Hermiticity:
\begin{align}
    i\frac{\partial}{\partial t}|\psi\rangle=H|\psi\rangle.\label{eq: non-hermitian-sch-eq}
\end{align}
This equation can describe lots of linear phenomena in various fields, including fluid mechanics \cite{Trefethen} and network science \cite{Asllani-2018}.
In such fields, $\ket{\psi}$ represents a linear deviation from the steady state whose norm can vary in time, and thus the norm is of interest in the relaxation process to the steady state.
We use the notation for the unnormalized time-evolved state $\ket{\psi(t)} = e^{-iHt}\ket{\psi(0)}$ and the normalized state $\ket{\tilde{\psi}(t)} = e^{-iHt}\ket{\psi(t)} / \| e^{-iHt}\ket{\psi(t)}\| $ in the following.

In the absence of non-normality, the norm dynamics is simply described by the imaginary part of the complex eigenenergies of the Hamiltonian. In most applications, the imaginary part is equal or less than zero for a stable steady state, and the largest value (the smallest absolute value) governs the long-time relaxation.

In the presence of non-normality, on the other hand, the dynamics is more complicated.
While the long-time behavior $(t\rightarrow\infty)$ is described by the imaginary part of the eigenspectrum of $H$ as in the absence of non-normality, the short-time behavior $(t\rightarrow0)$ is governed by the eigenspectrum of the Hermitian matrix $[H-H^\dagger]/i$ because
\begin{align}
    \langle\psi(t)|\psi(t)\rangle&=\langle\psi(0)|e^{iH^\dagger t}e^{-iHt}|\psi(0)\rangle\notag\\
    &\simeq1+t\langle\psi(0)|\frac{1}{i}[H-H^\dagger]|\psi(0)\rangle.
\label{eq: variation of norm}
\end{align}
Owing to this difference, the amplification of the norm is allowed in short time for positive eigenvalues of $[H-H^\dagger]/i$, even though the norm decays into zero in long time because of the non-positive imaginary part of the eigenspectrum of $H$.

The transient dynamics for intermediate values of $t$ ($0<t<\infty$) is more nontrivial, though it has often been overlooked \cite{Trefethen}.
From the third definition of the $\epsilon$-pseudospectrum, the state-vector corresponding to a pseudoeigenvalue is almost an eigenvector with a small deviation $\epsilon$.
If the imaginary part of the pseudoeigenvalue $a:=$Im$z$ is positive, there is a useful estimation for the lower bound of the norm amplification \cite{Trefethen}:
\begin{align}
    \sup_{0<t\leq\tau}\sqrt{\max_{\ket{\psi(0)}}\frac{\langle\psi(t)|\psi(t)\rangle}{\langle\psi(0)|\psi(0)\rangle}}\geq
    \frac{e^{a\tau}}{1+(e^{a\tau}-1)\epsilon/a}.
\end{align}
Roughly speaking, this bound implies that a pseudoeigenstate behaves as if it is an eigenstate for time $0<t<\tau$, where $\tau$ is estimated as
\begin{align}
    \frac{e^{a\tau}-1}{a}\sim\frac{1}{\epsilon}.\label{intermediate-time}
\end{align}

\begin{figure}[]
\begin{center}
 \includegraphics[width=6cm,angle=0,clip]{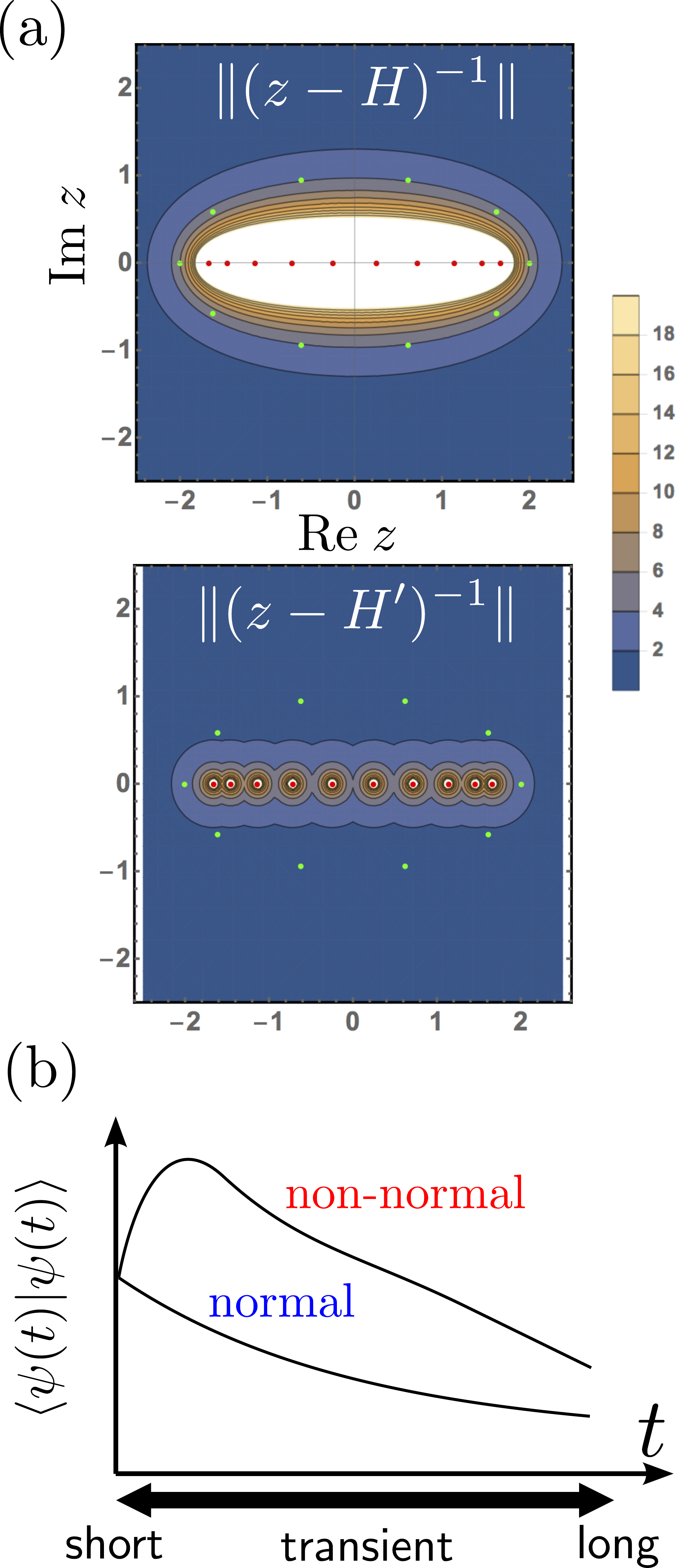}
 \caption{(a) Pseudospectra of OBC Hatano-Nelson model [$H$ in Eq.~\eqref{Hatano-Nelson}] and its Hermitian counterpart [$H'$ in Eq.~\eqref{similarity}] for $N=10, t=1,$ and $g=0.5$.
 We plot the value of $\|(z -A)^{-1}\|$ in the complex plane $z \in \mathbb{C}$ for $A=H,H'$. 
 One can see the $\epsilon$-pseudospectrum of $A$ as the region  $\{z\in\mathbb{C}~|~\|(z-A)^{-1}\|<1/\epsilon\}$.
 The green and red dots represent the PBC and the OBC spectra of the Hatano-Nelson model, respectively.
(b) Schematic picture of normal and non-normal norm dynamics.}
 \label{fig: schematic pictures}
\end{center}
\end{figure}

The norm dynamics with non-normal matrices discussed above is summarized in Fig. \ref{fig: schematic pictures}(b).
In each timescale, the dynamics is mainly governed by spectral, numerical, and pseudospectral abscissa defined as \cite{Trefethen}
\begin{align}
    \alpha(H)&=\sup \mathrm{Im}\sigma(H)~~~\mathrm{for~}t\rightarrow\infty,\label{spectralab}\\
    \omega(H)&=\sup \sigma\left(\frac{1}{2i}[H-H^\dagger]\right)~~~\mathrm{for~}t\rightarrow0,\label{numericalab}\\
    \alpha_{\epsilon}(H)&=\sup \mathrm{Im}\sigma_\epsilon(H)~~~\mathrm{for~}0<t<\infty,    \label{pspectralab}
\end{align}
respectively.
In the transient time-scale ($0<t<\infty$), one can consider arbitrary $\epsilon$ and $\alpha_\epsilon$, and the interplay among $\alpha(H), \omega(H), \alpha_\epsilon(H)$ in the dynamics is a matter of interest (see Ref.~\cite{Trefethen} for details).

\section{Non-normal Hamiltonian dynamics in quantum systems\label{sect: non-normal quantum dynamics}}
We here apply the mathematics of non-normal matrices reviewed in the previous section to non-Hermitian quantum systems defined as the continuous quantum trajectory of the Lindblad equation~\cite{Lindblad}.
We first characterize the probability for no quantum jump process by the non-normal norm dynamics.
We then consider the pseudospectral dynamics of a quantum state and derive the non-normal analog of the time-energy uncertainty relation.

\subsection{Non-normal norm dynamics in non-Hermitian quantum system \label{subsec: non-normal by Lindblad}}
In conventional quantum mechanics, Hamiltonians are assumed to be a Hermitian matrix/operator because of the unitarity of the quantum dynamics.
If one allows non-unitary processes such as dissipation and  measurements, however, one can consider a dynamics driven by a non-Hermitian Hamiltonian \cite{Daley-14,Nori-EP-19}.
Let us consider an open quantum system described by the Lindblad Master equation \cite{Lindblad}:
\begin{align}
    \frac{\partial \rho}{\partial t}&=-i[H_\mr{Herm},\rho]+\sum_{\mu}\left[\Gamma_\mu\rho\Gamma_\mu^\dagger-\left\{\frac{\Gamma_\mu^\dagger\Gamma_\mu}{2},\rho\right\}\right]\notag\\
    &=-i\left[H_{\rm eff}\rho-\rho H_{\rm eff}^\dagger\right]+\sum_{\mu}\Gamma_\mu\rho\Gamma_\mu^\dagger,\label{Lindblad}
\end{align}
where $\rho$ is the density matrix of the system, $H_\mr{Herm}$ is the Hermitian Hamiltonian that represents the unitary dynamics of the system, $\Gamma_\mu$'s are the jump operators that describe the system-environment coupling, and
\begin{align}
    H_{\rm eff}:= H_\mr{Herm}-\frac{i}{2}\sum_\mu\Gamma_\mu^\dagger\Gamma_\mu\label{effectiveham}
\end{align}
is the effective non-Hermitian Hamiltonian.
In the language of the quantum trajectory, the Lindblad dynamics of the density matrix is equivalent to the dynamics of the stochastic normalized pure state $|\tilde{\psi}(t)\rangle$
statistically-averaged over all trajectories for which the stochastic process during $t\rightarrow t+\delta t$ is defined as \cite{Daley-14}
\begin{itemize}
    \item With probability $\|e^{-iH_{\rm eff}\delta t}|\tilde{\psi}(t)\rangle\|^2$,
    \begin{align}
        |\tilde{\psi}(t+\delta t)\rangle:=\frac{e^{-iH_{\rm eff}\delta t}|\tilde{\psi}(t)\rangle}{\|e^{-iH_{\rm eff}\delta t}|\tilde{\psi}(t)\rangle\|}.\label{continuous}
    \end{align}
    \item With probability $\delta t~\|\Gamma_\mu|\tilde{\psi}(t)\rangle\|^2$,
    \begin{align}
        |\tilde{\psi}(t+\delta t)\rangle:=\frac{\Gamma_\mu|\tilde{\psi}(t)\rangle}{\|\Gamma_\mu|\tilde{\psi}(t)\rangle\|}.\label{jump}
    \end{align}
\end{itemize}
Equation (\ref{continuous}), which comes from the first term on the second line of Eq. (\ref{Lindblad}), describes the $continuous$ dynamics driven by the non-Hermitian effective Hamiltonian, while Eq. (\ref{jump}), which comes from the second term, does the $discontinuous$ quantum jump process. 
Note that the probability of no quantum jump is $\|e^{-iH_{\rm eff}\delta t}|\tilde{\psi}(t)\rangle\|^2 = 1 -i \delta t \expval{(H_\mr{eff} -H_\mr{eff}^\dag )}{\tilde{\psi}(t)} + O(\delta t^2) = 1 - \delta t \sum_\mu \expval{\Gamma_\mu^\dag \Gamma_\mu }{\tilde{\psi}(t)} + O(\delta t^2)$, so the sum of the probabilities for all the processes is unity.
If we focus on the trajectory with no quantum jump process, the whole dynamics of the stochastic wave function is completely captured by the non-Hermitian Schr\"{o}dinger equation (\ref{eq: non-hermitian-sch-eq}) except for the normalization of the state~\cite{Daley-14}.
In other words, when the stochastic dynamics is continuously monitored, one can extract the non-Hermitian Hamiltonian dynamics by post-selecting the continuous quantum trajectory.
Note that the amplification of the norm of the unnormalized state $\ket{\psi(t+\delta t)}$ is forbidden owing to the form of the effective non-Hermitian Hamiltonian \eqref{effectiveham}: it holds $1 - \delta t \sum_\mu \expval{\Gamma_\mu^\dag \Gamma_\mu }{\tilde{\psi}(t)} \leq 1$. The probability of the trajectory with no quantum jump stays or decays in time.
The decay of the norm of the unnormalized wave function corresponds to the disposal of the trajectories with quantum jumps.

Let us consider a probability to find a continuous trajectory with no quantum jump from $t'=0$ to $t'=t$.
By defining $N(t) = e^{-iH_\mr{eff} t}$, the probability of no quantum jump from $t$ to $t+\delta t$ is given by 
\begin{align}
    &p(t\rightarrow t+\delta t) =
    \norm\Big{ N(\delta t) \ket{\tilde{\psi}(t)} }^2 \notag \\
    &=
    \frac{\langle\tilde{\psi}(0)|N(t)^{\dagger}N(\delta t)^{\dagger}N(\delta t)N(t)|\tilde{\psi}(0)\rangle}{\langle\tilde{\psi}(0)|N(t)^{\dagger}N(t)|\tilde{\psi}(0)\rangle}\notag\\
    & =
    \left[\frac{\|N(t+\delta t)|\tilde{\psi}(0)\rangle\|}{\|N(t)|\tilde{\psi}(0)\rangle\|}\right]^2.
\end{align}
Therefore, the whole probability in the limit $\delta t\rightarrow0$ becomes
\begin{align}
    &P(t)=\prod_{t'=0,\delta t, \ldots, t-\delta t} p(t'\rightarrow t'+\delta t)\notag\\
    &\rightarrow
  \|e^{-itH_\mr{eff}}|\tilde{\psi}(0)\rangle\|^2,
\end{align}
i.e., the norm of the unnormalized wavefunction.

Thus far, we have discussed the quantum trajectory of the effective non-Hermitian Hamiltonian with the form~\eqref{effectiveham}.
For a given non-normal Hamiltonian $H_g$, it is possible to express $H_g$ in the form of Eq.~\eqref{effectiveham} as long as its non-Hermitian part $\frac{1}{2i}(H_g-H_g^\dag)$ is non-positive, i.e., $\omega(H_g) \leq 0$.
Namely, we can take $H_\mr{Herm} = \frac{1}{2}(H_g+H_g^\dag)$ and define $\Gamma_\mu$ by using, e.g., the spectral decomposition of the non-Hermitian part of $H_g$.
When the non-Hermitian part of $H_g$ is not non-positive, one can add a constant term to $H_g$ so that it becomes non-positive and the dynamics by $H_g$ is implementable as the Lindblad dynamics with no quantum jump.
The most ``efficient" choice of such constant is the numerical abscissa defined in Eq.~\eqref{numericalab},
\begin{equation}
 H_g \to H_g' = H_g - i \omega(H_g) \hat{1}, \label{eq: effective Ham}
\end{equation}
where $\hat{1}$ is the identity matrix.
The non-Hermitian part of $H_g'$ is apparently non-positive and the dynamics driven by $H_g'$ is realizable as the Lindblad equation.
With this shift, the probability to observe no quantum jump from $t'=0$ to $t'=t$ is given by
\begin{equation}
 P(t) = \|e^{-iH_g't} \ket{\tilde{\psi}(0)}\|^2 = e^{-2t\omega(H_g)} \|e^{-iH_gt} \ket{\tilde{\psi}(0)}\|^2.
 \label{success-rate}
\end{equation}
This is central to our discussion of the relationship between the probability of no quantum jump and the norm dynamics.
Note that any constant $c \geq \omega(H_g)$ suffices to make the non-Hermitian part of $H_g'$ non-positive.
Taking $c=\omega(H_g)$ is the most efficient in that the probability to observe no quantum jump becomes the largest.

The information of the norm dynamics and the numerical abscissa is included as the probability for the continuous quantum trajectory $P(t)$ [Eq.~\eqref{success-rate}].
For $\omega(H_g)=\alpha(H_g)$, the spectral abscissa (\ref{spectralab}), one can find an initial state such that $P(t)=1$ for any $t$, while for $\omega(H_g)>\alpha(H_g)$, any initial state decays in $t\rightarrow\infty$.
In the former case, the initial state is the eigenstate with no decoherence.
If one regards $\Gamma_{\mu}$ as a measurement,
this is a similar phenomenon to the quantum Zeno effect \cite{Zeno-77}, in which the dynamics of the two-level quantum system is frozen or suppressed by the continuous measurement.
The latter case can occur if and only if $H_g$ is non-normal.
Even in such a case, one can still consider large $P(t)$ for the transient time ($0<t<\infty$) if $\alpha_\epsilon(H_g)$, the pseudospectral abscissa (\ref{pspectralab}), is large enough. 
In this sense, one can also consider the non-normal analog of the quantum Zeno dynamics.
In the following, we consider the non-normal state dynamics in terms of the time-energy uncertainty relation.

\subsection{Non-normal quantum state dynamics: non-normal 
time-energy uncertainty relation \label{subsec: uncertainty relation}}
In the conventional quantum mechanics, the time-energy uncertainty relation for a quantum state $|\psi\rangle$ is
given by \cite{Jammer}
\begin{align}
    \Delta E\Delta t\sim1,\label{uncertainty}
\end{align}
where 
\begin{align}
    \Delta E:=\sqrt{\langle \psi|H^2|\psi\rangle-(\langle \psi|H|\psi\rangle)^2}\label{convfluctuation}
\end{align}
is the fluctuation of the energy
and $\Delta t$ is the time in which the overlap $\langle \psi(0)|\psi(t)\rangle=\langle \psi|e^{-iHt}|\psi\rangle$ is not small. 

In principle, it is possible to generalize the time-energy uncertainty relation to a normalized quantum state in non-Hermitian quantum mechanics.
We here discuss the uncertainty relation for a pseudoeigenstate with pseudoeigenenergy $E_\epsilon$ with large imaginary part of a normal/non-normal Hamiltonian $H$.
Instead of the conventional fluctuation (\ref{convfluctuation}), we define 
\begin{align}
    \Delta E_{\epsilon}:=\mathrm{Im~} E_\epsilon-\sup \mathrm{Im~}\sigma(H)=\mathrm{Im~} E_\epsilon-\alpha(H).\label{unconvfluctuation}
\end{align}
Since the non-normal state dynamics in a non-Hermitian quantum system is determined by the non-Hermitian Schr\"{o}dinger equation (\ref{eq: non-hermitian-sch-eq}) except for the normalization factor, one can estimate the time in which the overlap between the normalized states, $\langle \tilde{\psi}(0)|\tilde{\psi}(t)\rangle$, becomes small by using the non-normal matrix theory introduced in Sect. \ref{sect: basics-of-non-normal}.
In particular, the definition of the transient time (\ref{intermediate-time}), in which a pseudoeigenvector behaves as if it is an eigenvector, can be regarded as a generalization of the time fluctuation in time-energy uncertainty relation.
By setting $a=\Delta E_{\epsilon}$ and $\tau=\Delta t$ in Eq. (\ref{intermediate-time}), we obtain
\begin{align}
    \Delta E_{\epsilon}\Delta t\sim\log\left[1+\frac{\Delta E_{\epsilon}}{\epsilon} \right].\label{n-uncertainty}
\end{align}
For a non-normal Hamiltonian, $\Delta E_{\epsilon}$ can be much larger than $\epsilon$, and the right-hand side can take a large value. 
Equivalently, the fluctuation $\Delta t$, in which the pseudoeigenvector behaves like an eigenvector, is much larger than the inverse of the fluctuation $\Delta E_{\epsilon}$ in the presence of large non-normality. 
For example, $\Delta E_{\epsilon}/\epsilon$ is exponentially large with respect to the system size in the Hatano-Nelson model, as mentioned in the previous section.

Note that Eq. (\ref{n-uncertainty}) reproduces the conventional time-energy uncertainty relation (\ref{uncertainty}) for a Hermitian Hamiltonian ($\Delta E_{\epsilon}\sim\epsilon$), which is a consequence of the rough equivalence between the definitions (\ref{convfluctuation}) and (\ref{unconvfluctuation}) \footnote{Suppose that $\Delta E_{\epsilon}=E_{\epsilon}/i=\epsilon$. Then for a unit pseudoeigenvector, $H|v\rangle=i\epsilon|v\rangle+\ket{\mathcal{O}(\epsilon)}$, where $\ket{\mathcal{O}(\epsilon)}$ is a vector whose norm is $\mathcal{O}(\epsilon)$. Since $\langle v|\mathcal{O}(\epsilon)\rangle$ is $\mathcal{O}(\epsilon)$, Eq.(\ref{convfluctuation}) is shown to be $\mathcal{O}(\epsilon)$. }.

\section{Non-Hermitian Hamiltonian dynamics on quantum computers \label{sect: theory of quantum circuits}}
In this section, we describe theories for realizing the non-normal Hamiltonian dynamics on quantum computers~\cite{Nielsen2011}.
First, we discuss the direct implementation of the Lindblad equation on a non-unitary quantum circuit with a continuous measurement.
As another approach, we next consider a quantum-classical hybrid algorithm called variational quantum simulation (VQS)~\cite{Li2017efficient,Endo2020general}.
In the first approach, we can directly simulate the Lindblad equation, but the probability for obtaining the trajectory with no quantum jump decays 
with the simulation time as shown in Eq.~\eqref{success-rate}.
The second approach can treat the norm directly in the simulation, but it assumes a trial functional form of wavefunctions (called {\it ansatz}) under the time evolution and the result of the dynamics can be affected by the choice of the ansatz.

\subsection{Direct realization of Lindblad dynamics as a non-unitary circuit \label{subsec: non-unitary implemtation}}
Terashima and Ueda~\cite{Terashima-Ueda-03} have shown that if ancilla qubits are available, the controlled-NOT gate, all one-qubit unitary gates, and the one-qubit projective measurement constitute a universal set for a non-unitary quantum circuit $N$ (a non-unitary matrix of the size $2^n \times 2^n$ for $n$ qubit systems).
They have used the singular value decomposition of a matrix $N=UDV$, where $U$ and $V$ are unitary matrices and $D$ is a diagonal matrix.
The unitaries $U$ and $V$ can be decomposed into CNOT gates and one-qubit unitary gates~\cite{Nielsen2011}, and they have proved that $D$ can be decomposed into a product 
of one-qubit non-unitary gates with $n-1$ control qubits, which is realized with a projective measurement on one or two ancilla qubits in a probabilistic way.
Note that the ancilla qubits required here can be {\it recycled} for sequential applications of the one-qubit non-unitary gates with $n-1$ control qubits, so it is enough to have one or two ancilla qubits during the operation.  
In the framework of the quantum measurement, the implemented non-unitary operation in the original system (without ancilla qubits) is seen as one of the measurement (Kraus) operators $M_{\mu}$ that satisfy $\sum_{\mu}M_{\mu}^\dagger M_{\mu}=1$.
When we denote $M_0$ as the desired operation and redefine $M_1$ such that it represents the other operations, Ueda and Terashima have showed that they are expressed as
\begin{align}
    M_0&=c~\frac{N}{\max_{\ket{\tilde{\phi}}} \|N|\tilde{\phi}\rangle\|}, \label{eq: original M0} \\
    M_1&=\sqrt{1-M_0^\dagger M_0},
\end{align}
where $0< c\leq1$ is a constant depending on the implementation.
The $success$ rate of the desired operation for a quantum state $|\tilde{\psi}\rangle$ is given by
\begin{align}
    p=\langle \tilde{\psi}|M_0^\dagger M_0|\tilde{\psi}\rangle=
    |c|^2\left[\frac{\|N|\tilde{\psi}\rangle\|}{\max_{\ket{\tilde{\phi}}} \|N|\tilde{\phi}\rangle\|} \right]^2\leq1.
\end{align}

Now we are in a position to implement the quantum trajectory driven by a Hamiltonian $H_g$ on a non-unitary quantum circuit.
For simplicity, we assume $c=1$.
Under this assumption, the Hamiltonian dynamics is characterized by the non-unitary operation (measurement)
\begin{align}
    M_0 = \frac{N(\delta t)}{\max_{\ket{\tilde{\phi}}} \|N(\delta t)|\tilde{\phi}\rangle\|},\label{short-time-operation}
\end{align}
where $N(t) = e^{-iH_g t}$ as in the previous section.
In the non-unitary quantum circuit, the set of ancilla qubits can be seen to play a role of the environment.
In this interpretation, the whole Lindblad dynamics is implemented just by repeating this measurement.
If we post-select the trajectory in which the measurement $M_0$ is successful at all time (or at all repetitions), the initial state of the system obeys the quantum dynamics described by the effective Hamiltonian~\eqref{eq: effective Ham}. As we discussed, the probability for having this trajectory is given by Eq.~\eqref{success-rate} (recall that $\max_{\ket{\tilde{\phi}}} \|N(\delta t)|\tilde{\phi}\rangle\| = 1 + 2\delta t \cdot \omega(H_g) + O(\delta t^2)$).

Note that if we regard the set of the measured ancilla qubits as a {\it system} conversely and consider the situation where they are measured only at once during a time step $t\rightarrow t+\delta t$ and reused in all time steps, the dynamics of the ancilla qubits mimics the setup of the quantum Zeno effect~\cite{Zeno-77} coupled with {\it an environment} (the original $n$ qubit system in the Lindblad equation).
An implementation satisfying the assumption is always possible, though it require more ancilla qubits than those in the minimum set because we cannot recycle the ancilla qubits in the same time step.

Finally, we discuss how we utilize the direct implementation of the Lindblad equation discussed above to simulate dynamics under a given $n$-qubit non-normal Hamiltonian $H_g$ in practice.
When $n$ is large, it is classically impossible to perform the singular value decomposition of the $2^n \times 2^n$ matrix $N(\delta t)$.
In that case we may use the Suziki-Trotter decomposition of $N(\delta t)$ as long as $H_g$ is written as the small number (e.g. poly($n$)) of Pauli matrices, $H_g = \sum_i c_i P_i + i \sum_i d_i Q_i$, where $c_i, d_i \in \mathbb{R}$ and $P_i$ and $Q_i$ are Pauli matrices on $n$ qubits.
Namely, we approximate $N(\delta t)$ as $N(\delta t) \approx \Pi_i e^{-i\delta t c_i P_i} \dot \Pi_i e^{\delta t d_i Q_i}$ and consider implementation of the non-unitary gate $e^{\delta t d_i Q_i}$.
Since $Q_i$ is a Pauli matrix, it is possible to transform it to a single qubit $Z$ gate with a unitary $u$ composed of CNOT gates and $H, S^\dag$ gates~\cite{Nielsen2011}, $e^{\delta t d_i Q_i} = u^\dag e^{\delta t d_i Z} u$, and the implementation of a single qubit non-unitary gate $e^{\delta t d_i Z}$ can be done with the method by Terashima and Ueda.
The normalization factor (denominator) in Eq.~\eqref{eq: original M0} for $N=e^{\delta t d_i Z}$ is $1+\delta t |d_i|$,  so the total factor for $\Pi_i e^{\delta t d_i Q_i}$ is $1 + \delta t \sum_i |d_i|$.
This means that the probability $P(t)$ [Eq.~\eqref{success-rate}] becomes $P(t) = e^{-2t\sum_i |d_i|} \|e^{-iH_gt} \ket{\tilde{\psi}(0)}\|$, which is less efficient than when we directly implement $e^{-iH_gt}$ because $\omega(H_g) = \|\frac{1}{2i}(H_g-H_g^\dag) \| = \| \sum_i d_i Q_i \| \leq \sum_i |d_i|$.
Note that the same decomposition for non-unitary part of the time evolution operator was recently used in a proposal of the digital quantum simulation of the Lindblad dynamics~\cite{kamakari2021digital}, although the non-unitary gate $e^{\delta t d_i Q_i}$ was implemented in a different way.

\subsection{Realization of non-normal dynamics by variational quantum simulation \label{subsec: VQS implement}}
\begin{figure*}[t]
  \centering
  \mbox{
  \Qcircuit @C=2.0em @R=0.7em {
  \lstick{\ket{0}_{\mr{anc}}} & \qw & \qw & \qw & \gate{H} &  \ctrl{2}  & \gate{X} &  \qw & \qw & \ctrl{2} & \gate{H} &  \meter\\
  &&\cdots&&&&\cdots&&&&\\
  \lstick{\ket{\phi_0}} & {/}_{n} \qw & \gate{U_1} & \qw & \gate{U_{i-1}} & \gate{P_i} &  \gate{U_i} & \qw & \gate{U_{j-1}} & \gate{P_j} &  \qw & \qw \\
  }
  }
  \\ \vspace{5mm}
  \mbox{
  \Qcircuit @C=2.0em @R=0.7em {
  \lstick{\ket{0}_{\mr{anc}}} & \qw & \qw & \qw & \gate{H} &  \ctrl{2}  & \qw &  \qw & \qw & \ctrl{2} & \gate{S} & \gate{H} & \meter\\
  &&\cdots&&&&\cdots&&&&\\
  \lstick{\ket{\phi_0}} & {/}_{n} \qw & \gate{U_1} & \qw & \gate{U_{j-1}} & \gate{P_{j}} &  \gate{U_j} & \qw& \gate{U_{K}} & \gate{G_m} &  \qw & \qw \\
  }
  }
  \caption{
    (top) Quantum circuit for evaluating $M_{i,j}$ ($i<j$ and $i,j \neq 0$).
    The expectation value of the measurement on an ancilla qubit yields $M_{i,j}/(\alpha^2 \cdot g_i g_j)$.
    When $i=0$ and $j\neq 0$, we have $M_{0,j} = 0$ (see the main text).
    For $i=j$, $M_{i,j}$ is exactly known as 1 $(i=0)$ and $g_i^2 \alpha^2$ ($i\neq 0$).
    (bottom) Quantum circuit for evaluating $V_j$ ($j \neq 0$).
    Here we decompose the Hamiltonian into a sum of unitaries $\{G_m\}_m$ (e.g., Pauli matrices) as $ H^\dag = \sum_m c_m G_m, c_m \in \mathbb{C}$. The presented circuit calculates a contribution from $G_m$; 
    the expectation value of the measurement yields
    $ \alpha^{-2} \cdot \mr{Im} \qty( \bra{\Psi(\vec{\Theta}(t))} G_m \pdv{\ket{\Psi(\vec{\Theta}(t))}}{\Theta_j} )$.
    For $j=0$, we have $V_0 = \alpha \cdot \mr{Im} \expval{H^\dag}{\tilde{\Psi}(\theta_1,\ldots,\theta_K)}$, which can be evaluated by standard expectation value measurements for $G_m$ with the normalized state $\ket{\tilde{\Psi}(\theta_1,\ldots,\theta_K)}$.
  } 
  \label{fig: circuitMV}
\end{figure*}
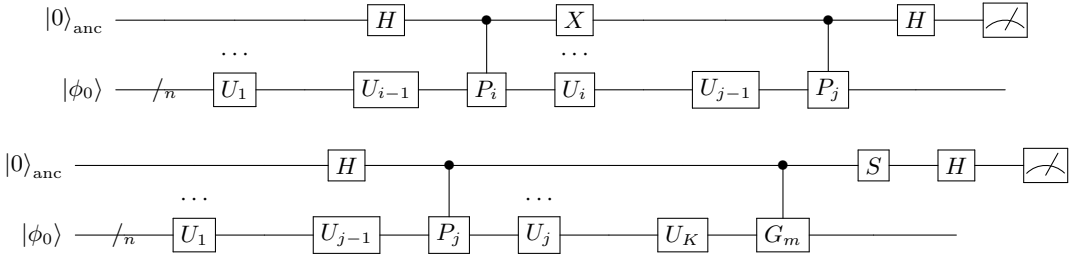

As another approach to simulate non-normal dynamics of quantum systems on quantum computer, we adopt variational quantum simulation (VQS)~\cite{Li2017efficient, Endo2020general}.
VQS is a quantum-classical hybrid algorithm to simulate the dynamics of quantum states.
A parametrized quantum state called {\it ansatz} state is chosen, and its parameters are updated to reproduce the dynamics of the target system by employing the information of measurements on quantum computers.
The first proposal of VQS treats real-time evolution of pure states under Hermitian Hamiltonian, but it is extended to general Hamiltonian including non-normal ones and general Lindblad time-evolution of mixed states~\cite{Endo2020general}.
In the following, we review the VQS algorithm with remarking several points peculiar to the non-normal Hamiltonian dynamics.

Let us consider an $n$-qubit quantum state, an {\it ansatz} state, defined as follows:
\begin{align}
\ket{\Psi (\alpha, \theta_1, \ldots, \theta_K)} = \alpha \cdot U_{K}(\theta_{K}) \cdots U_{1}(\theta_1)\ket{\phi_0},
\label{eq: ansatz}
\end{align}
where $\alpha \in \mathbb{R}$ is a parameter which we call norm parameter, $U_i(\theta_i)$ is a unitary gate with parameter $\theta_i$, $K$ is the number of parameters, and $\ket{\phi_0}$ is a reference state to create the ansatz state.
Note that this state is not normalized in general: $\braket{\Psi(\alpha, \theta_1, \ldots, \theta_K)|\Psi(\alpha, \theta_1, \ldots, \theta_K)}=\alpha^2$.
We assume that $U_i(\theta_i)$ is composed of a non-parametrized gate $R_i$ and a single Pauli rotation gate $e^{i g_i \theta_i P_i}$ with a coefficient $g_i \in \mathbb{R}$ and a $n$-qubit Pauli matrix $P_i \in \{I,X,Y,Z \}^{\otimes n}$, i.e., $U_i(\theta_i) = R_i e^{ig_i P_i}$.
We shall introduce the notations $\vec{\theta} :=(\theta_1, \ldots, \theta_K), \vec{\Theta} := (\Theta_0, \ldots, \Theta_K) := (\alpha, \theta_1,\ldots, \theta_K)$ and the normalized state
$\ket{\tilde{\Psi}(\theta_1,\ldots, \theta_K)} := U_{K}(\theta_{K}) \cdots U_{1}(\theta_1)\ket{\phi_0}$ for simplicity.

When initial parameters $\alpha(0), \vec{\theta}(0)$ (i.e., an initial state $\ket{\Psi(\alpha(0), \vec{\theta}(0))}$) and a non-normal Hamiltonian $H$ are given, the VQS algorithm calculates the dynamics of the parameters $\alpha(t), \vec{\theta}(t)$ that approximates the solution of the Schr{\"o}dinger equation, 
\begin{equation}
 \dv{\ket{\psi}}{t} = - iH \ket{\psi},
 \label{eq: original schroedinger}
\end{equation}
within the Hilbert space spanned by the ansatz state, $\{ \ket{\Psi(\alpha, \vec{\theta})} \}_{\alpha, \vec{\theta}}$.
Namely, Eq.~\eqref{eq: original schroedinger} is mapped to the time evolution of the ansatz parameters $\alpha(t), \theta(t)$ by minimizing the distance between the exact time-evolved states under Eq.~\eqref{eq: original schroedinger} and the ansatz state under infinitesimal time-variation $\delta t$~\cite{Li2017efficient}, 
\begin{align}
 \min_{\dot{\alpha}, \dot{\vec{\theta}}} \, \delta \left\lVert \qty( \dv{t} + iH ) \ket{\Psi (\alpha(t), \vec{\theta}(t))} \right\rVert, 
\end{align}
where $\|\ket{\varphi} \|=\sqrt{\braket{\varphi|\varphi}}$ is the norm of $\ket{\varphi}$.
Minimization can be exactly performed when ignoring $O(\delta t^2)$ and higher-order terms, resulting in
\begin{align}
 \sum_{j=0}^K M_{i,j} \dot{\Theta}_j = V_i
 \label{eq: MthetaV}
\end{align}
for $i=0,\ldots,K$, where $\dot{\Theta}_j = \dv{\Theta_j}{t}$ and
\begin{equation}
 \begin{aligned}
	  M_{i,j} & = \mr{Re} \qty(\pdv{ \bra{\Psi(\vec{\Theta}(t))}}{ \Theta_i} \pdv{\ket{\Psi(\vec{\Theta}(t))}}{\Theta_j} ),\\
      V_j & = -\mr{Im} \qty( \bra{\Psi(\vec{\Theta}(t))} H^{\dag}  \pdv{ \ket{\Psi(\vec{\Theta}(t))} }{\Theta_j} ).
 \label{eq: MV}
 \end{aligned}
\end{equation}
Those matrix $M$ and vector $V$ can be measured on quantum computers of $n+1$ qubits by running quantum circuits depicted in Fig.~\ref{fig: circuitMV}, as proposed in Refs.~\cite{Li2017efficient, Endo2020general}.
Note that the derivatives with respect to $\alpha$ are evaluated in slightly different way as those with respect to the other parameters $\vec{\theta}$.

To simulate the dynamics of the ansatz parameters, one first calculates the time-derivative of $\vec{\Theta}(t)$ at time $t$ by evaluating Eq.~\eqref{eq: MV} using quantum computers and solving Eq.~\eqref{eq: MthetaV}.
Then the value of $\dot{\vec{\Theta}}$ is utilized to obtain the time-evolved parameters $\vec{\Theta}(t+\delta t)$:
for example, $\vec{\Theta}(t+\delta t) = \vec{\Theta}(t+\delta t) + \delta t \cdot \dot{\vec{\Theta}}(t)$ in the Euler method.
Compared with the conventional approach to simulate the dynamics of quantum systems by performing the Suzuki-Trotter decomposition to the time-evolution operator, $e^{iHt} \approx (e^{iH\delta t})^{t/\delta t}$, VQS is more friendly to the hardware of quantum computers in that the depth of quantum circuits used in the algorithm is fixed and does not grow in time.
Although it is not evident what ans{\"a}zte should be used and not analytically analyzable how the choice of them affects the result of the simulation {\it a priori}, VQS stands as a promising approach to simulate the general dynamics in quantum systems on current quantum computers without error correction~\cite{Preskill2018}.

Lastly, we comment several points specific to VQS of the non-normal dynamics.
First, because of $H \neq H^\dag$, there appears $H^\dag$ in the right hand side of the definition $V$ [Eq.~\eqref{eq: MV}].
Second, when we adopt the ansatz including the norm parameter $\alpha$ like Eq.~\eqref{eq: ansatz}, the coefficient matrix $M$ of Eq.~\eqref{eq: MV} becomes block-diagonal,
\begin{align}
 \begin{pmatrix}
 1 & 0 & \cdots & 0 \\
 0 & \tilde{M}_{1,1} & \cdots & \tilde{M}_{1,K}\\
 \vdots & \vdots & \ddots & \vdots \\
 0 & \tilde{M}_{K,1} & \cdots & \tilde{M}_{K,K}& 
 \end{pmatrix}
 \begin{pmatrix}
  \dot{\alpha} \\ \dot{\theta}_1 \\ \vdots \\ \dot{\theta}_K
 \end{pmatrix}
 =
 \begin{pmatrix}
 V_0 \\ \tilde{V}_1 \\ \vdots \\ \tilde{V}_K
 \end{pmatrix},
\end{align}
where
\begin{align}
\begin{split}
 \tilde{M}_{i,j}
= \mr{Re} \qty(\pdv{ \bra{\tilde{\Psi}(\vec{\theta})}}{ \theta_i} \pdv{\ket{\tilde{\Psi}(\vec{\theta})}}{\theta_j} ), \\
\tilde{V}_j = - \mr{Im} \qty( \bra{\tilde{\Psi}(\vec{\theta})} H^\dag \pdv{\ket{\tilde{\Psi}(\vec{\theta})}}{\theta_j} ),
\end{split}
\end{align}
for $i,j=1,\ldots,K$.
Particularly, the time derivative of the norm parameter $\dot{\alpha}$ is described as
\begin{align}
\dot{\alpha} &= \alpha \cdot \mr{Im} \expval{H^{\dag}}{\tilde{\Psi}(\vec{\theta})}
\\
&= -\alpha \cdot \frac{1}{2i} \expval{(H-H^\dag)}{\tilde{\Psi}(\vec{\theta})}
\end{align}
This is consistent with Eq.~\eqref{eq: variation of norm}.

\section{Experiments on quantum hardware\label{sect: experiment}}
In this section, we present experimental results of the non-normal Hamiltonian dynamics simulation realized on cloud-based quantum computers provided by IBM Quantum~\cite{IBMQ2021}.
We consider the Hatano-Nelson model~\eqref{Hatano-Nelson} of two-sites (where there is no different between the PBC and OBC).
To make the non-normality large in this very small model, we set $t=g=1/2$ so that the Hamiltonian becomes non-diagonalizable (called {\it exceptional} point): 
\begin{align}
    H=
    \begin{pmatrix}
    0&0\\
    1&0
    \end{pmatrix}.
    \label{exceptional}
\end{align}
We assume the standard basis $|0\rangle:=(1,0)^T,|1\rangle:=(0,1)^T$.
The time evolution operator $N(t)$ in this case is exactly calculated as
\begin{equation}
 N(t) = e^{-iHt} =
 \begin{pmatrix}
  1 & 0 \\ -it & 1
 \end{pmatrix}.
\end{equation}
We consider three initial states for the dynamics:
\begin{equation}
\ket{\tilde{\psi}(0)} = \ket{1}, \frac{1}{\sqrt{2}}(\ket{0} - i\ket{1}), \ket{0}.   
\end{equation}
We focus on two quantities during the dynamics: the probability to obtain trajectory with no quantum jump, $P(t)$ [Eq.~\eqref{success-rate}], and the overlap between the normalized time-evolved state and the initial state, $L(t) = |\braket{\tilde{\psi}(t)|\tilde{\psi}(0)}|$.
The first one reflects information of the spectrum of the non-normal Hamiltonian $H$ as discussed in Sect.~\ref{subsec: non-normal by Lindblad} and the second one is related to the time-energy uncertainty relation in Sect.~\ref{subsec: uncertainty relation}.

In the following, we present experimental results of the non-normal Hamiltonian dynamics by Eq. \eqref{exceptional} implemented in two ways introduced in the previous section:
the non-unitary quantum circuit and VQS.

\subsection{Non-unitary quantum circuit}
\begin{figure}
    \begin{minipage}{\columnwidth}
        \mbox{
        \Qcircuit @C=1em @R=.7em {
        \lstick{\ket{0}_{\rm sys}} & \gate{U_{\rm init}} & \gate{V}& \ctrl{1}          & \gate{U} & \gate{U_{\rm init}^\dag}&\meter\\
        \lstick{\ket{0}_{\rm anc}} & \qw                 &   \qw   & \gate{U(A_-/A_+)} & \meter   & \qw                     &\qw
        \gategroup{1}{3}{2}{5}{.7em}{--}      \\
        \\
        &&& \mathrm{repeat}~t/\delta t~\mathrm{times}&&
        \\
        }
        }
    \end{minipage}\\
    \caption{Non-unitary quantum circuit for non-normal Hamiltonian (\ref{exceptional}). The non-unitary circuit consists of one system qubit for the non-unitary evolution and one ancilla qubit for the measurement.
    The non-unitary operation (\ref{eq: t->t+dt}) at each time step $t\rightarrow t+\delta t$ is defined in the dotted box. 
    The norm dynamics is obtained as the success probability for the non-unitary circuit, while the state dynamics is observed as the overlap between the initial state $\ket{\tilde{\psi}(0)} = U_{\rm init}\ket{0}$ and $|\tilde{\psi}(t)\rangle$.
    }\label{fig: non-unitary-circuit}
\end{figure}
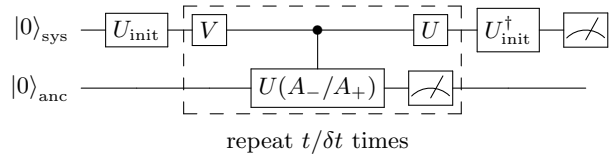

As we described in Sect.~\ref{subsec: non-unitary implemtation}, any non-unitary operation can be implemented by using the singular value decomposition and projective measurements on ancilla qubits.
By using this technique, we decompose the short-time non-unitary operation (\ref{short-time-operation}) defined for the Hamiltonian (\ref{exceptional}):
\begin{align}
    M_0&=\frac{1}{A_{+}}
    \begin{pmatrix}
        1&0\\
        -i \cdot \delta t&1
    \end{pmatrix}\notag\\
    &=U
    \begin{pmatrix}
        1&0\\
        0&A_{-}/A_{+}
    \end{pmatrix}V,\label{eq: t->t+dt}
\end{align}
where $A_{\pm}=\sqrt{2+\delta t^2\pm \delta t\sqrt{4+\delta t^2}}/\sqrt{2}$ are the singular values of $e^{-iH\delta t}$.
This choice of $M_0$ corresponds to $c=1$ in Eq.~\eqref{eq: original M0} and the effective Hamiltonian describing trajectory with no quantum jump is $H' = H - i/2 \hat{1}$ because $\omega(H) = 1/2$. 
Reference~\cite{Terashima-Ueda-03} has constructed a non-unitary gate for a diagonal matrix $N(a)=$diag $(1,a)$ by using a controlled-unitary gate that consists of a system qubit $|\psi\rangle_{\rm sys}$
and an initialized ancilla qubit $|0\rangle_{\rm anc}$.
In this construction, the unitary matrix
\begin{align}
    U(a)=
    \begin{pmatrix}
    a&\sqrt{1-a^2}\\
    \sqrt{1-a^2}&-a
    \end{pmatrix}
\end{align}
is acted on the ancilla qubit controlled by the system qubit, and then the projective measurement on the ancilla qubit is performed.
If the ancilla qubit stays in $|0\rangle_{\rm anc}$ after the measurement, the non-unitary operation $N(a)$ on the system qubit is successful. 
By using this construction, we obtain the circuit representation for the non-unitary dynamics by the Hamiltonian \eqref{exceptional} [Fig. \ref{fig: non-unitary-circuit}].
In this circuit, the continuous Hamiltonian dynamics can be detected by post-selecting the measurement outcomes in which the ancilla qubit stays in $|0\rangle_{\rm anc}$.
Without post-selection, the whole Lindblad dynamics is recovered.

We use a cloud-based quantum computer \verb|ibmq_athens| provided by IBM Quantum~\cite{IBMQ2021}.
We set relatively large time interval $\delta t=1$ to reduce the number of repetitions of the time evolution $N(\delta t)$ and not to make the fidelity of the whole quantum circuit so small.
We run the quantum circuit in Fig.~\ref{fig: non-unitary-circuit} for 4096 times and evaluate $P(t)$ by counting the number of runs where the measurement result ``0"  in the ancilla qubit is observed for all time steps.
The overlap $L(t)$ is determined as the probability for obtaining ``0" in the system qubit after applying $U_\mr{init}^\dag$ of the time-evolved state, $|\braket{0|U_\mr{init}^\dag|\tilde{\psi}(t)}|^2$, where $U_\mr{init}$ is a unitary that produces an initial state as $\ket{\tilde{\psi}(0)}=U_\mr{init}\ket{0}$.
We use 4096 shots to determine $L(t)$.

\begin{figure*}
\begin{center}
    \includegraphics[width=0.45\textwidth]{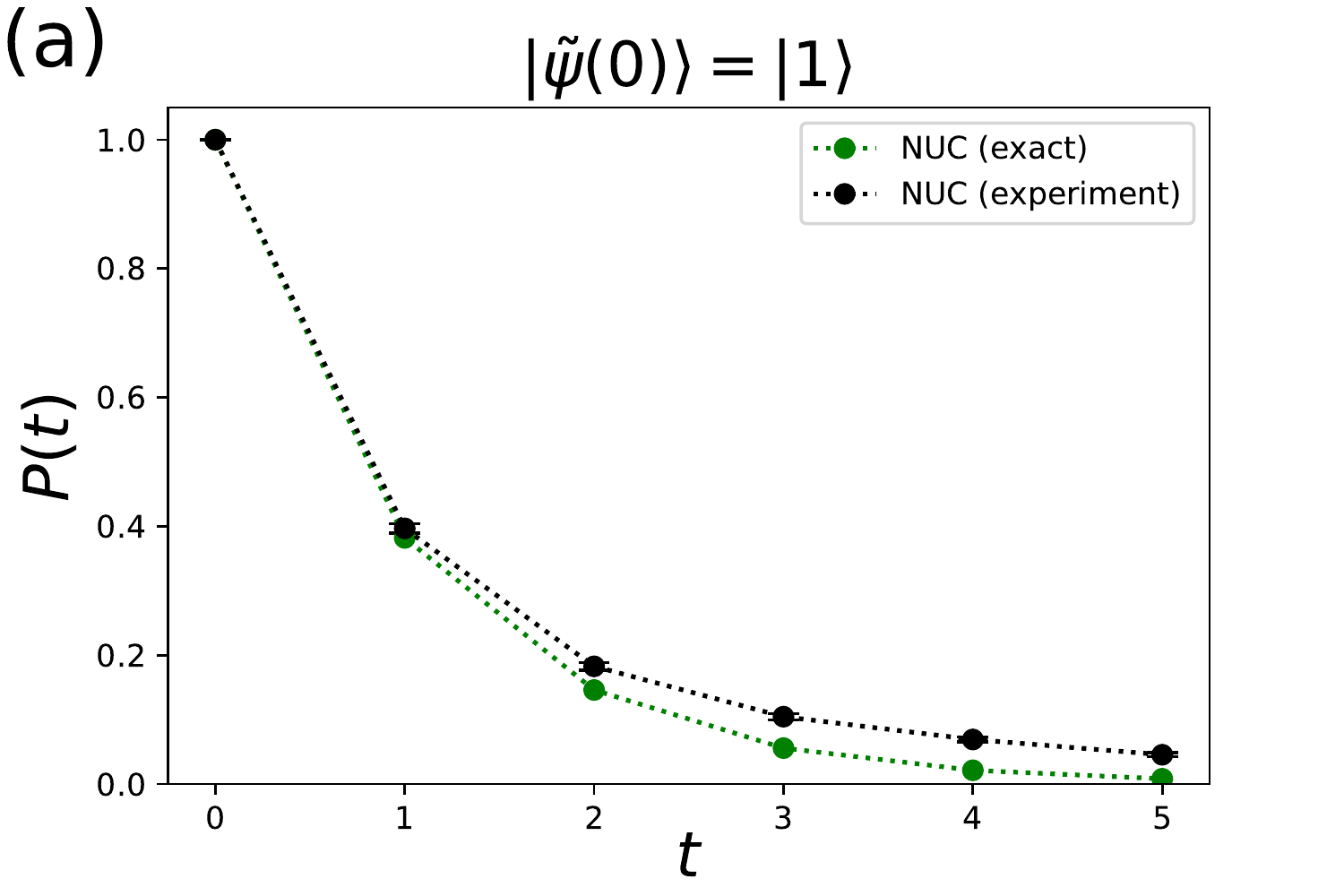}
    \includegraphics[width=0.45\textwidth]{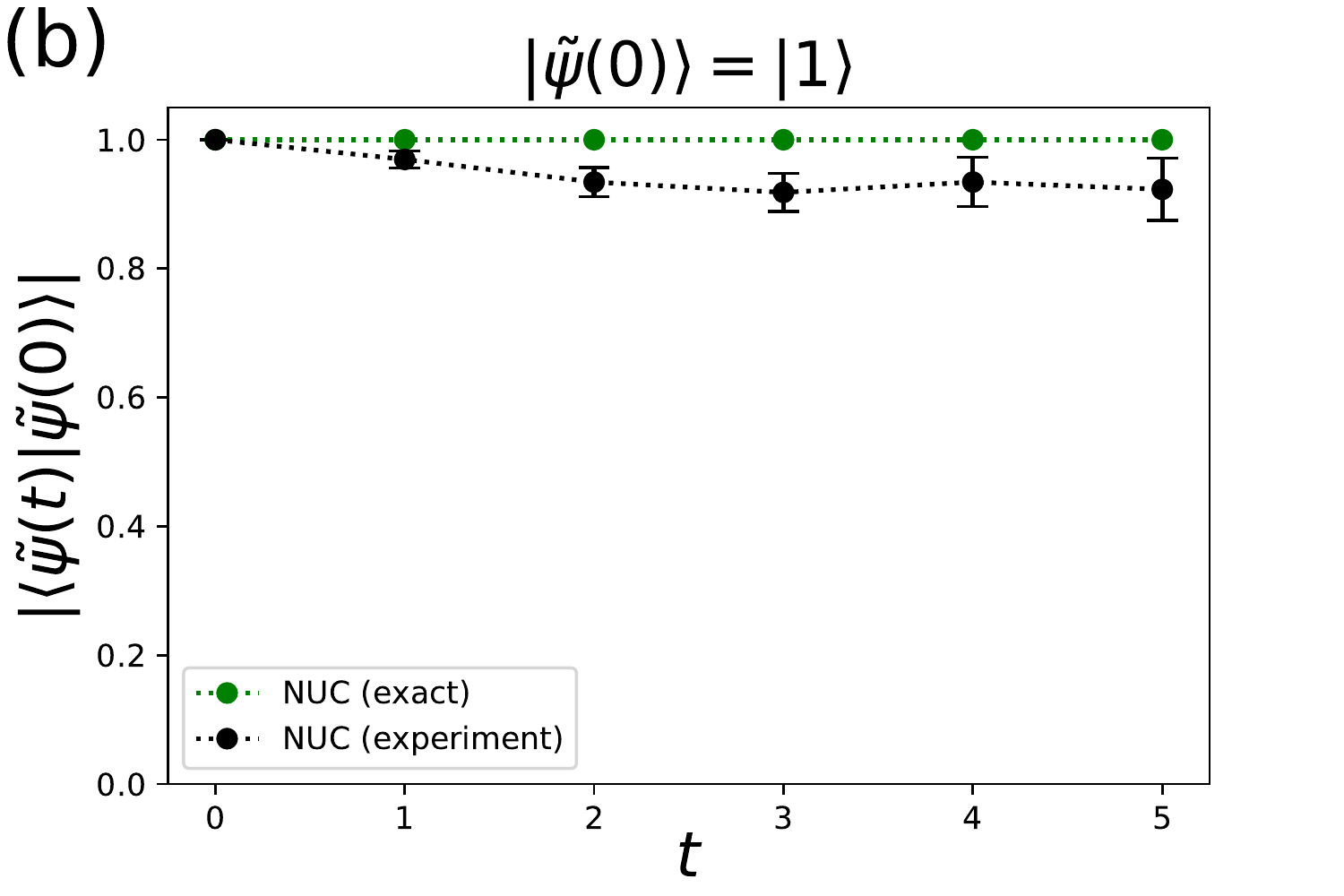}
    \includegraphics[width=0.45\textwidth]{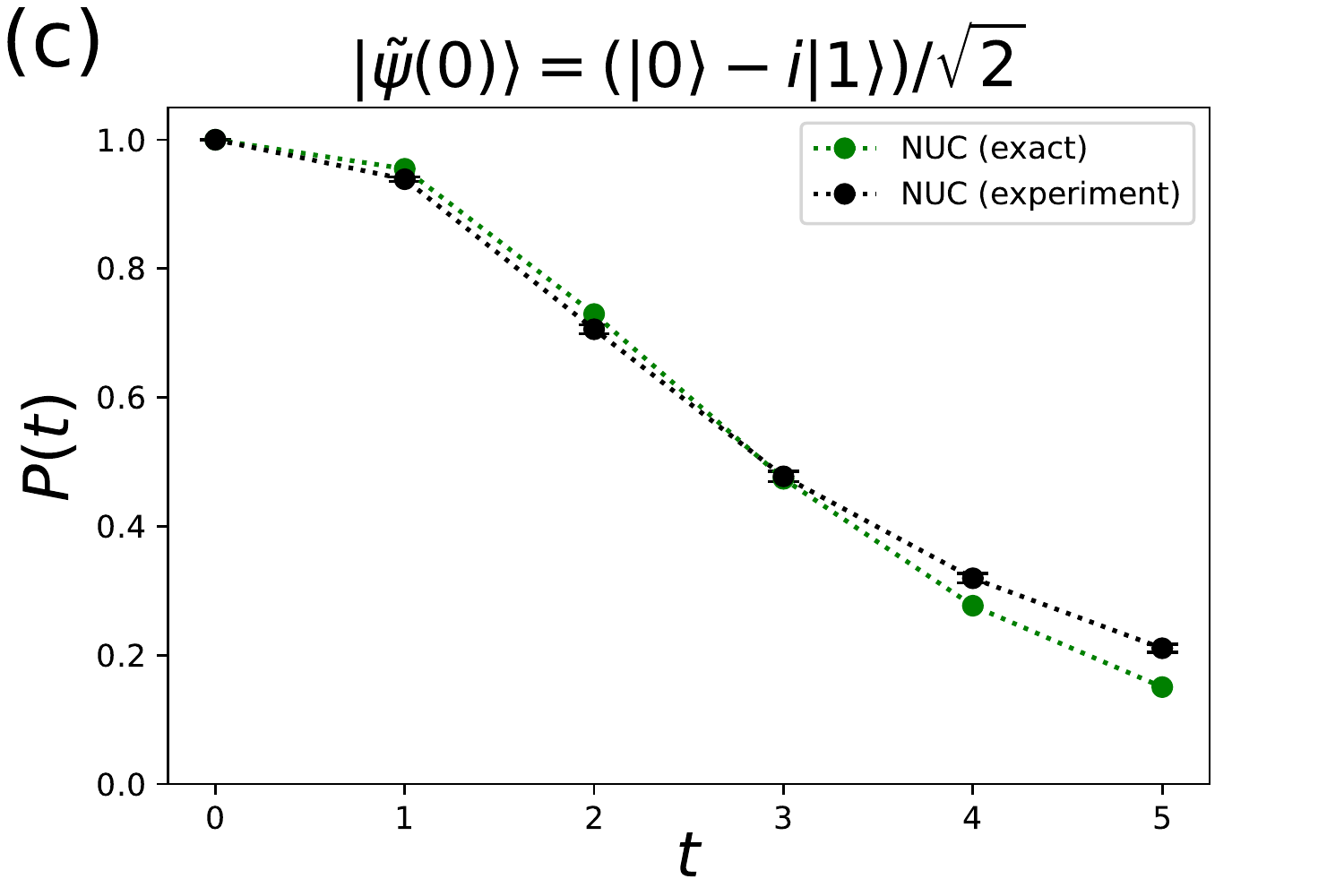}
    \includegraphics[width=0.45\textwidth]{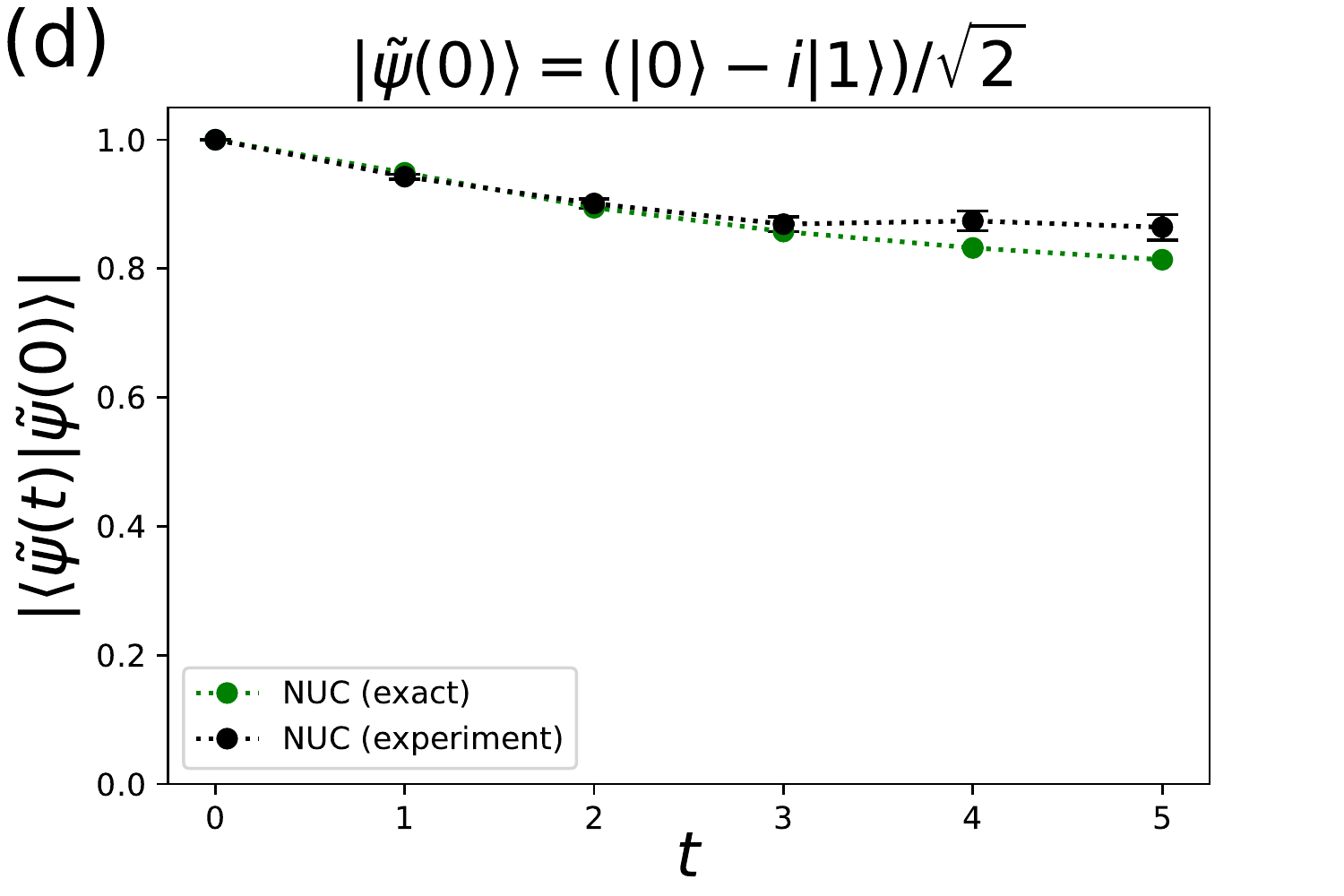}
    \includegraphics[width=0.45\textwidth]{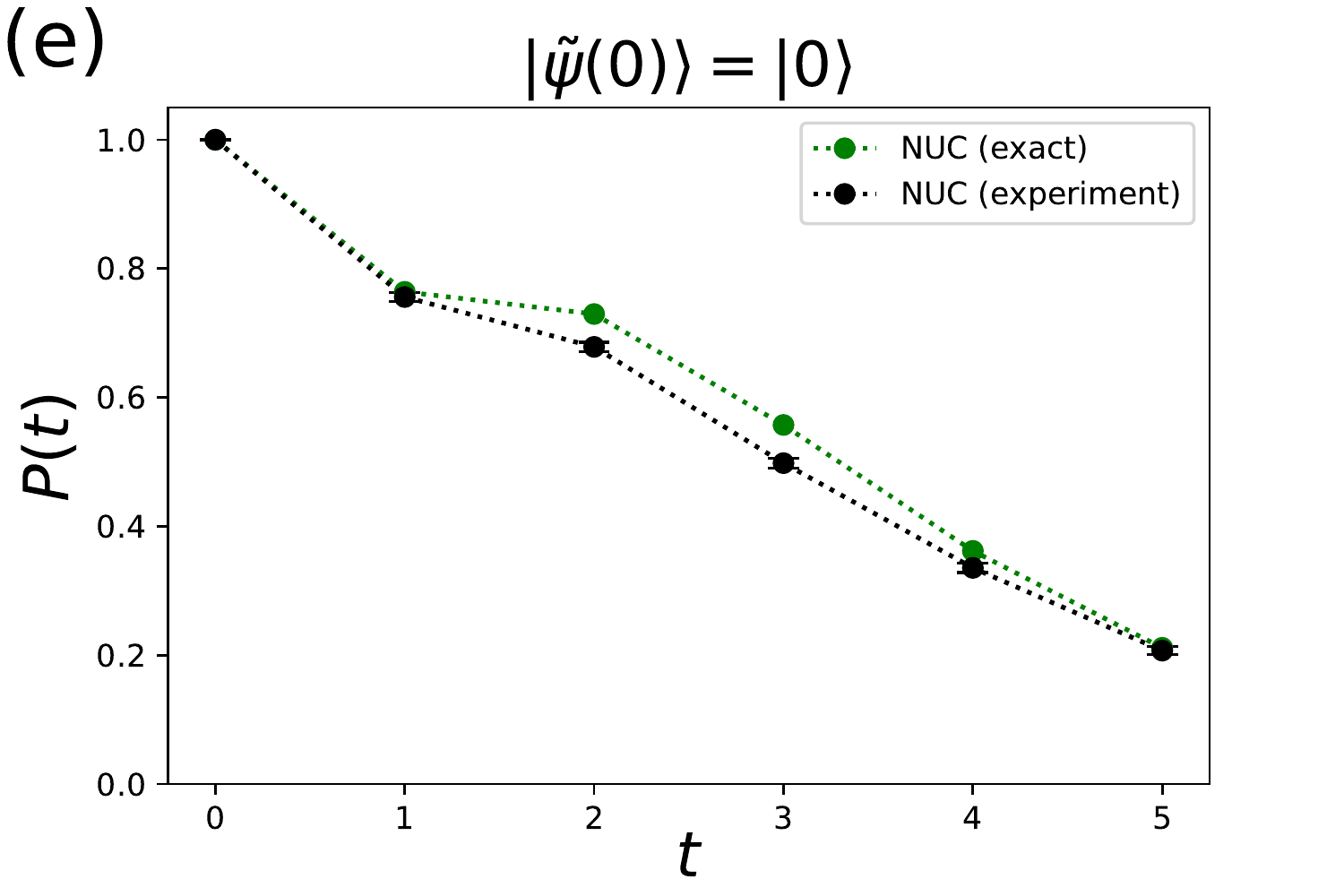}
    \includegraphics[width=0.45\textwidth]{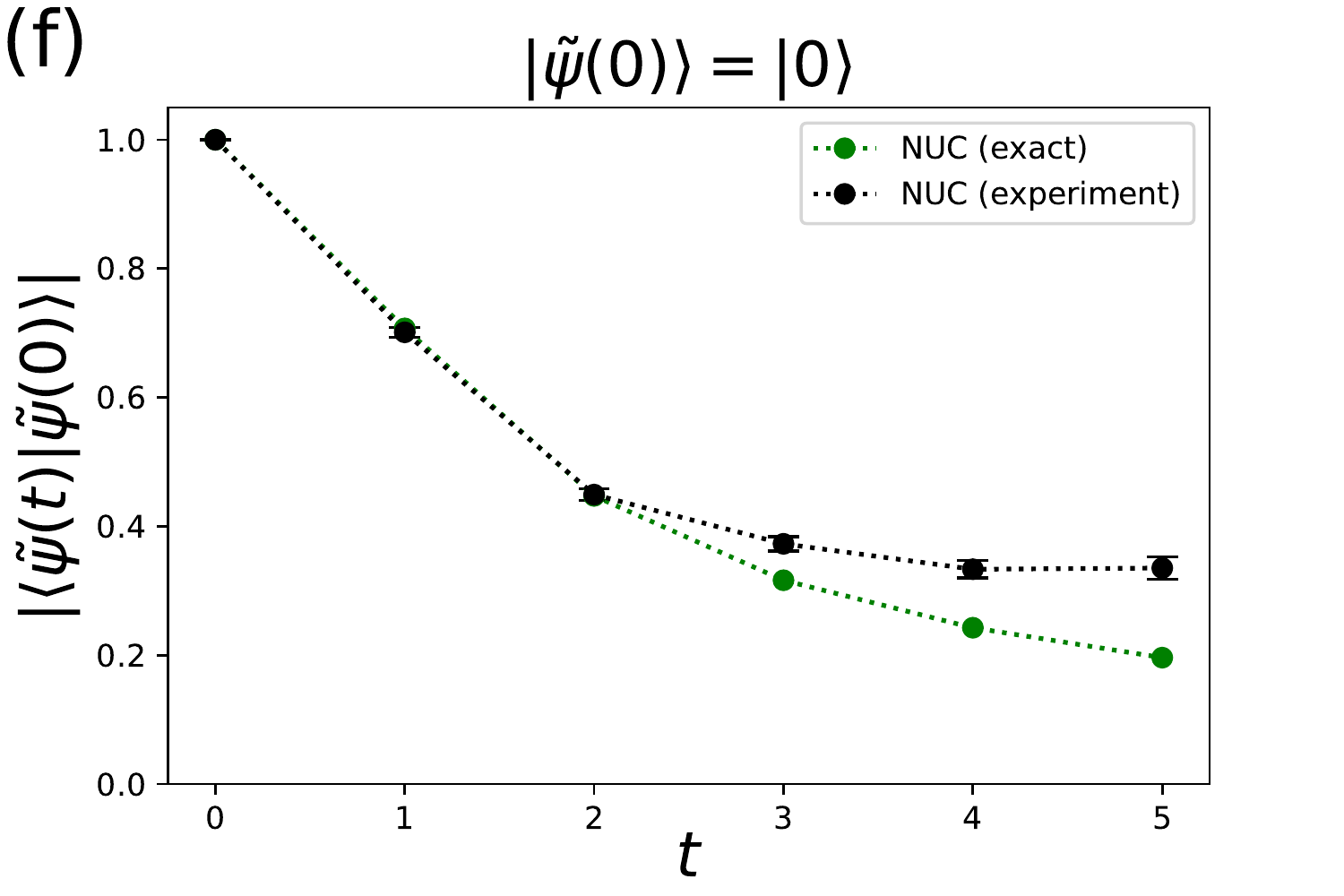}
 \caption{Result of running non-unitary quantum circuit (NUC) on real quantum hardware.
    (a,c,e) Dynamics of the success probability for no quantum jump, $P(t)$.
    (b,d,f) Dynamics of the overlap between the normalized state and the initial state, $|\ip{\tilde{\psi}(t)}{\psi_0}|$.
    ``NUC (exact)" indicates the dynamics obtained by a numerical simulation on a classical computer.
    When there is no noise and error in the experiment, ``NUC (exact)" matches with the experimental value ``NUC (experiment)".
    The error bars are estimated as the standard deviation of the binomial distribution for 1024 shots.}
 \label{fig: NUC results}
\end{center}
\end{figure*}

The result is shown in Fig.~\ref{fig: NUC results}.
For the initial state $|\tilde{\psi}(0)\rangle=|1\rangle$, $P(t)$ exponentially decays in time as in the case of the normal dynamics because it is the eigenstate of the effective Hamiltonian $H' =H - i/2\hat{1}$. 
Ideally, the overlap $L(t)$ becomes unity for the eigenstate, but the small deviation is observed because of the noise and error in the real quantum hardware.
For the initial state $|\tilde{\psi}(0)\rangle=(|0\rangle-i|1\rangle)/\sqrt{2}$, the dynamics reflects the non-normal nature.
Since this initial state is not the eigenstate of $H'$ with real energy, the probability $P(t)$ should decay in the long-time limit $t\rightarrow\infty$.
However, it is the eigenstate of $(H-H^{\dagger})/2i$ with the largest eigenvalue and satisfies
\begin{align}
    \left.\frac{dP(t)}{dt}\right|_{t=0}=0,
\end{align}
which leads to the small decay rate of $P(t)$ for short and transient time thanks to the non-normality.
If we focus on the ancilla qubit, this dynamics can be seen as a non-normal analog of the quantum Zeno effect in the open quantum system.
We see this behavior of $P(t)$ in the experimental results in Fig.~\ref{fig: NUC results} (middle row).
Moreover, $L(t)$ stays at large value for transient time even though the initial state is not an eigenvalue of $H'$.
This is a consequence of the pseudospectral nature of the initial state, or the non-normal time-energy-uncertainty relation (\ref{n-uncertainty}). 
Finally, for the initial state $\ket{\tilde{\psi}(0)}=\ket{0}$, we see the decay of $P(t)$ {\it and} $L(t)$ although there is some plateau in $P(t)$ in transient time.
This is because the initial state is neither the eingenstate of $H'$ or $\frac{1}{2i}(H - H^\dag)$, so there is no general reason for $P(t)$ or $L(t)$ to get large in any time scale.

Note that most of the data points obtained from the actual quantum computer in Fig.~\ref{fig: NUC results} deviate the ones from the exact calculation though their behaviors are qualitatively the same.
The deviation comes from the noise in the gate operations and the measurement error in the current quantum hardware.

\subsection{Variational quantum simulation}
Next, we provide an experimental demonstration of the non-normal Hamiltonian dynamics on real quantum devices by using VQS explained in Sect.~\ref{subsec: VQS implement}.
We employ the ansatz state defined as
\begin{equation}
\ket{\Psi(\alpha, \theta)} = \alpha \cdot e^{-\frac{i}{2} \theta X} \ket{0}.
\end{equation}
The initial parameters corresponding to the initial states $\ket{\tilde{\psi}(0)} = \ket{1}, \frac{1}{\sqrt{2}}(\ket{0} - i\ket{1})$, and $\ket{0}$ are $(\alpha(0), \theta(0)) = (1, \pi), (1, \pi/2)$ and $(1, 0)$, respectively.
Under this ansatz, Eq.~\eqref{eq: MthetaV} takes the following form:
\begin{equation}
 \begin{pmatrix}
 1 & 0 \\ 0 & \frac{\alpha^2}{4}
 \end{pmatrix}
 \begin{pmatrix}
 \dot{\alpha} \\ \dot{\theta}
 \end{pmatrix}
 =
 \begin{pmatrix}
 - \alpha g \expval{Y}{\tilde{\Psi}(\theta)} \\ 
 \frac{\alpha^2}{2} (1 + g\expval{Z}{\tilde{\Psi}(\theta)})
 \end{pmatrix},
 \label{eq: MV RX ansatz}
\end{equation}
where $\ket{\tilde{\Psi}(\theta)} = e^{-\frac{i}{2} \theta X} \ket{0}$ is the normalized state realizable on quantum computers.
Note that we do not need any ancilla qubit to evaluate the above quantities because the ansatz contains only one rotational parameter and quantum circuits like Fig.~\ref{fig: circuitMV} get simple.
Namely, the right hand side of Eq.~\eqref{eq: MV RX ansatz} can be evaluated by expectation values of Pauli operator $Y$ and $Z$ of a single-qubit state $\ket{\tilde{\Psi}(\theta)}$.

We use cloud-based quantum computers \verb|ibmq_belem| and \verb|ibmq_quito| provided by IBM Quantum~\cite{IBMQ2021} to run VQS experimentally.
The expectation values $\expval{Y}{\tilde{\psi}(\theta)}$ and $\expval{Z}{\tilde{\psi}(\theta)}$ are measured by 1024 shots, and the obtained values are fed into the VQS algorithm to calculate the dynamics of the ansatz parameters $(\alpha(t), \theta(t))$.
We calculate the overlap $L(t) = |\braket{\tilde{\psi}(t)|\tilde{\psi}(0)}|$ by using the relation $|\braket{\tilde{\psi}(t)|\tilde{\psi}(0)}| = \alpha |\expval{e^{i\frac{\theta(t)}{2}X} \cdot e^{-i\frac{\theta(0)}{2}X}}{0}|$.
That is, $L(t)$ is evaluated by the probability of measuring $\ket{0}$ for the state $e^{i\frac{\theta(t)}{2}X} e^{-i\frac{\theta(0)}{2}X} \ket{0}$ by 1024 shots.
To mitigate the measurement error~\cite{Qiskit}, we evaluate the probability matrix
\begin{equation}
 P_{\mr{prob}} =
 \begin{pmatrix}
  p_{0|0} & p_{0|1} \\ p_{1|0} & p_{1|1}
 \end{pmatrix},
\end{equation}
where $p_{i|j}$ is a probability of measuring $i$ when the state is $\ket{j}$ ($i,j=0,1$). 
We performed the evaluation of the matrix beforehand the VQS experiments with using 1024 shots for each component of the matrix.
The measurement count of 0 (1), denoted as $C_0 (C_1)$, obtained in the VQS experiments is modified as
\begin{equation}
 \begin{pmatrix} \tilde{C}_0 \\ \tilde{C}_1 \end{pmatrix}
 = P_{\mr{prob}}^{-1}
 \begin{pmatrix} C_0 \\ C_1 \end{pmatrix}.
\end{equation}

\begin{figure*}
    \centering
    \includegraphics[width=0.45\textwidth]{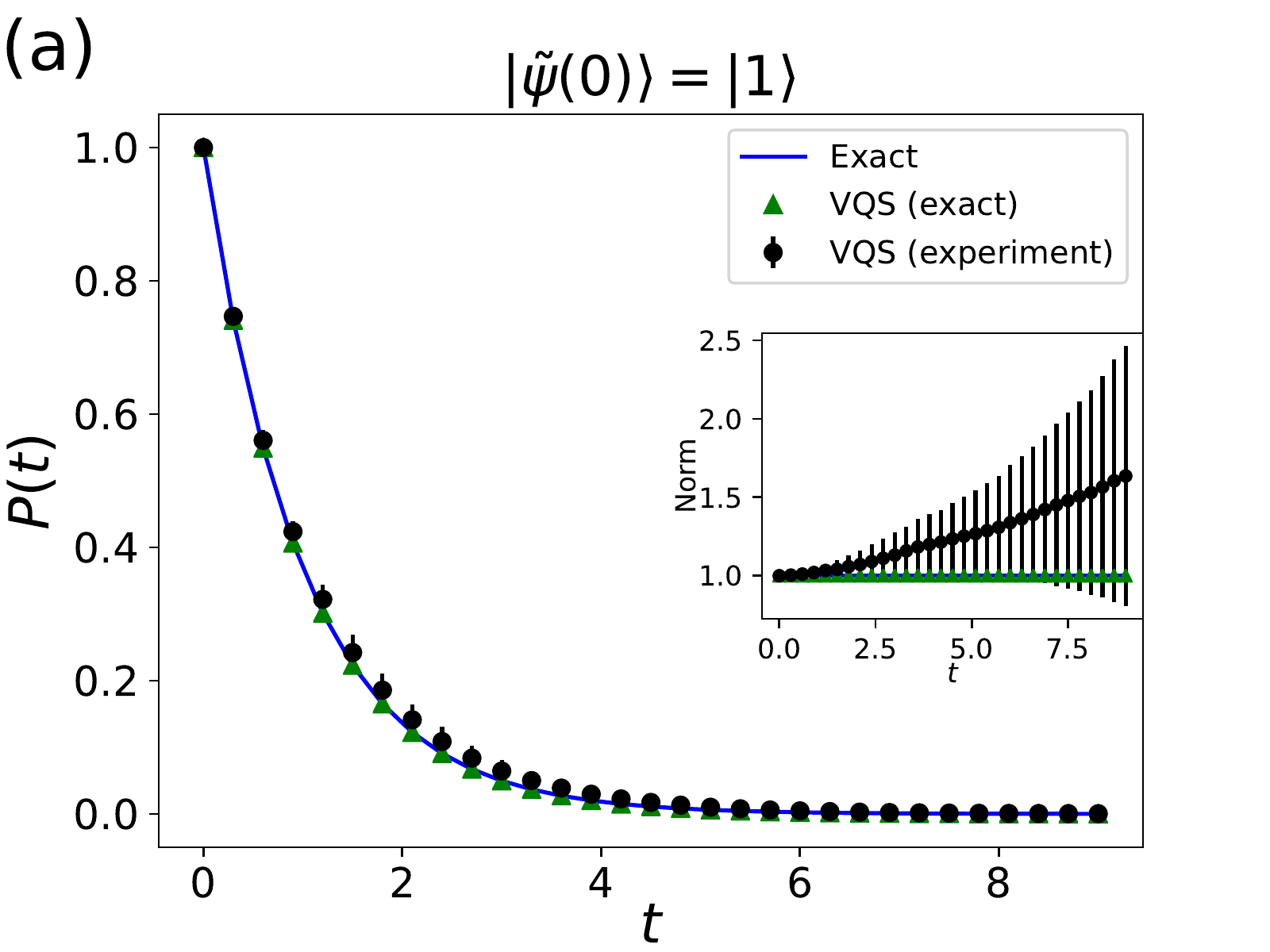}
    \includegraphics[width=0.45\textwidth]{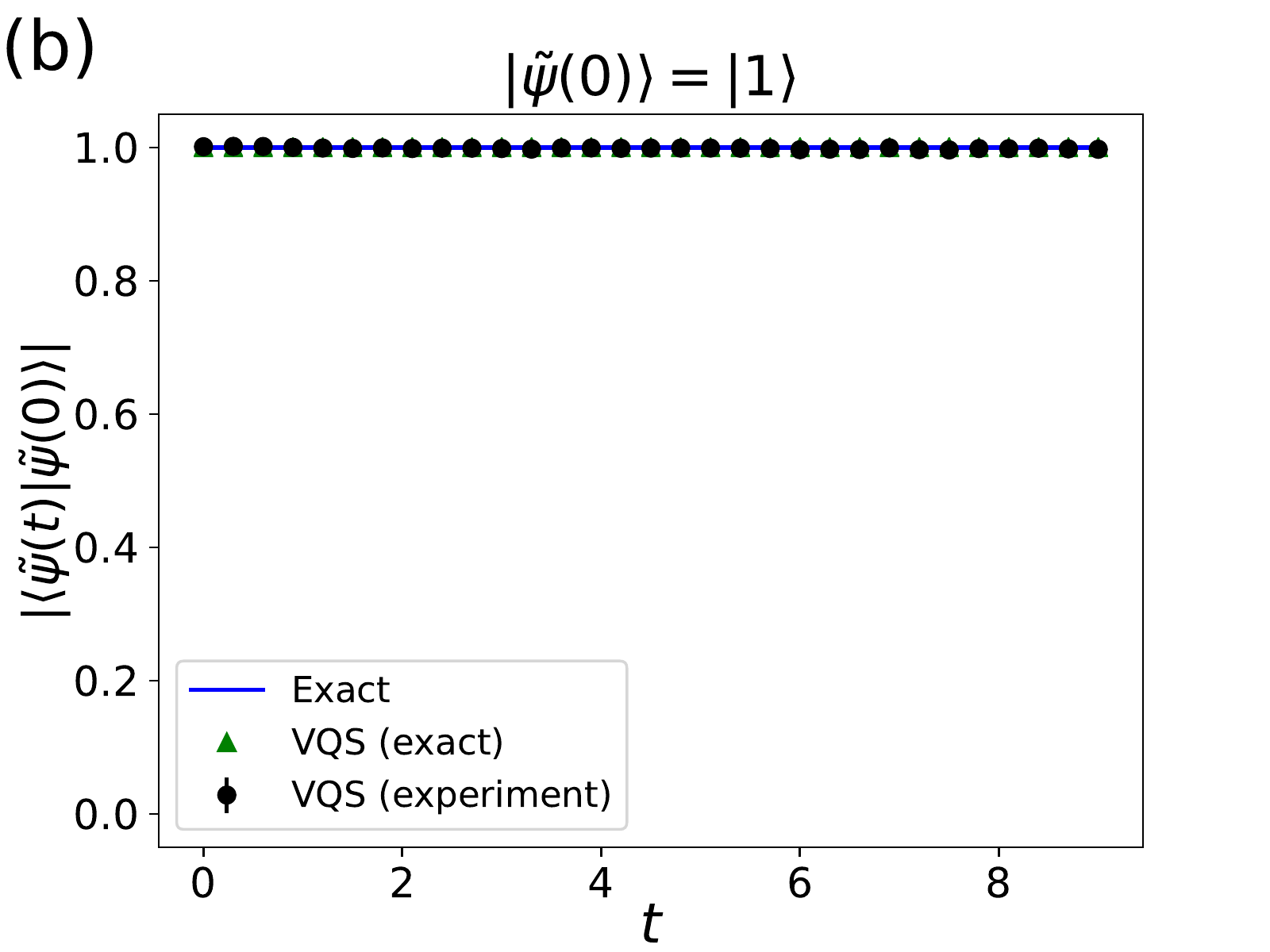}
    \includegraphics[width=0.45\textwidth]{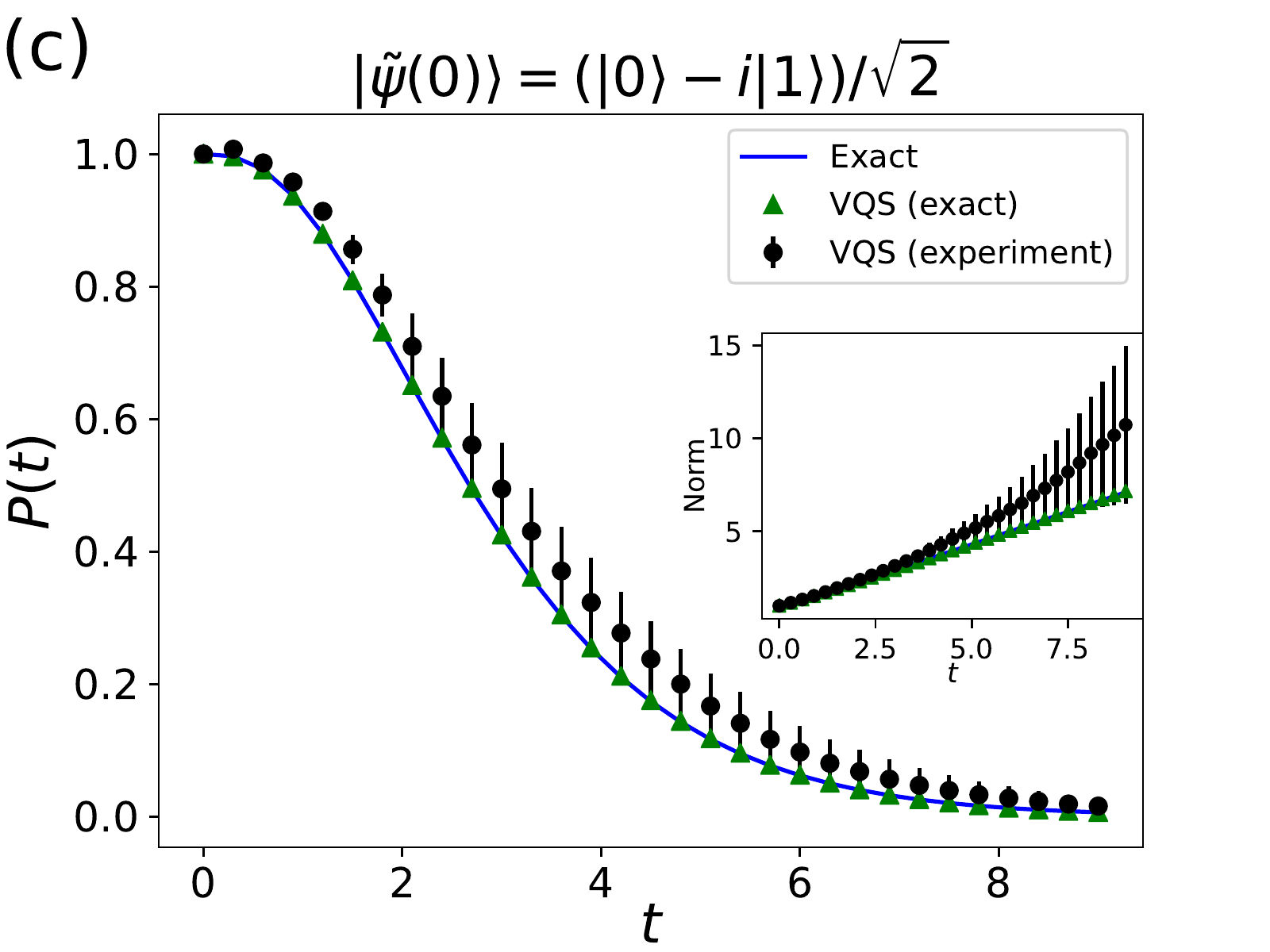}
    \includegraphics[width=0.45\textwidth]{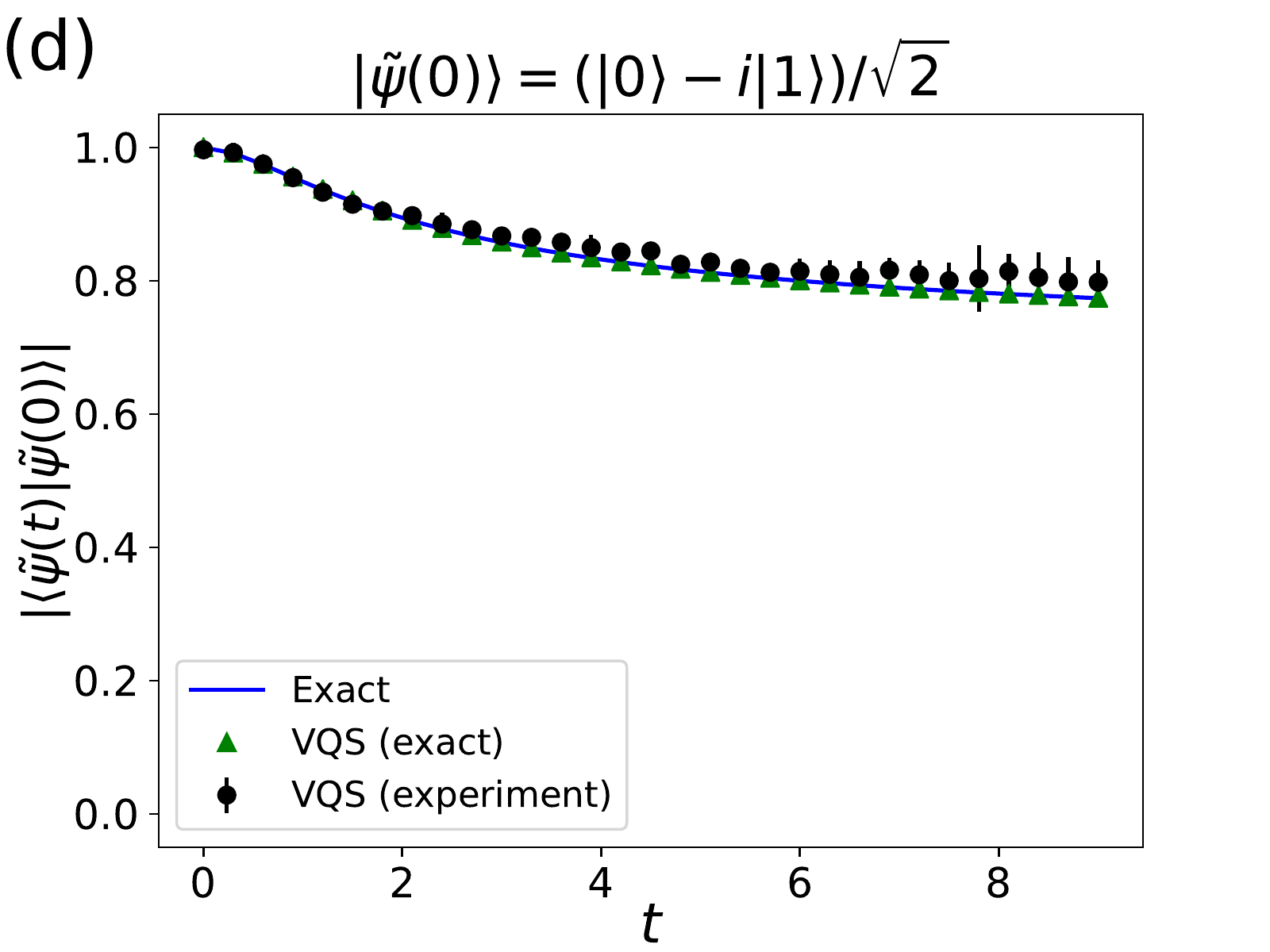}
    \includegraphics[width=0.45\textwidth]{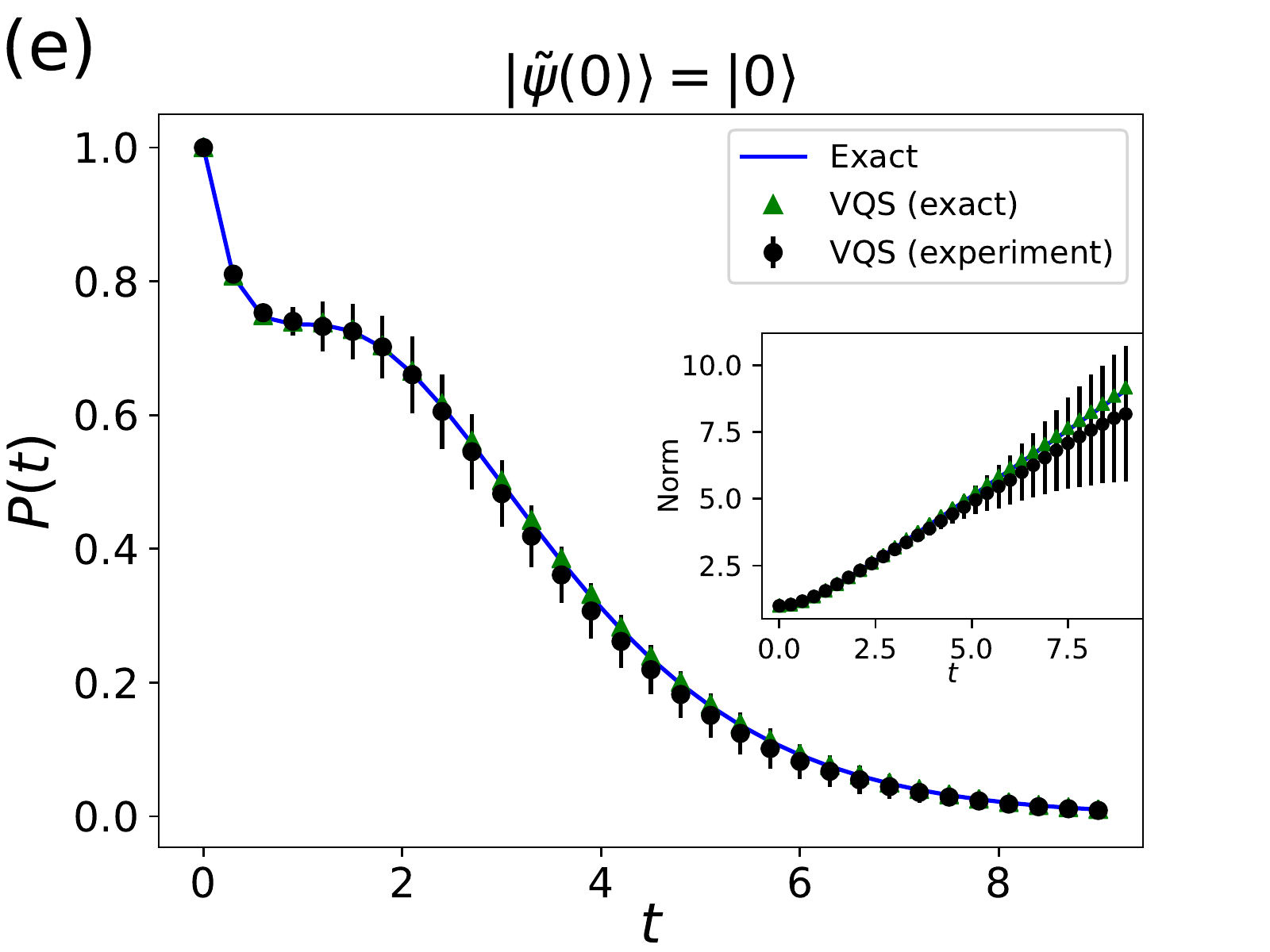}
    \includegraphics[width=0.45\textwidth]{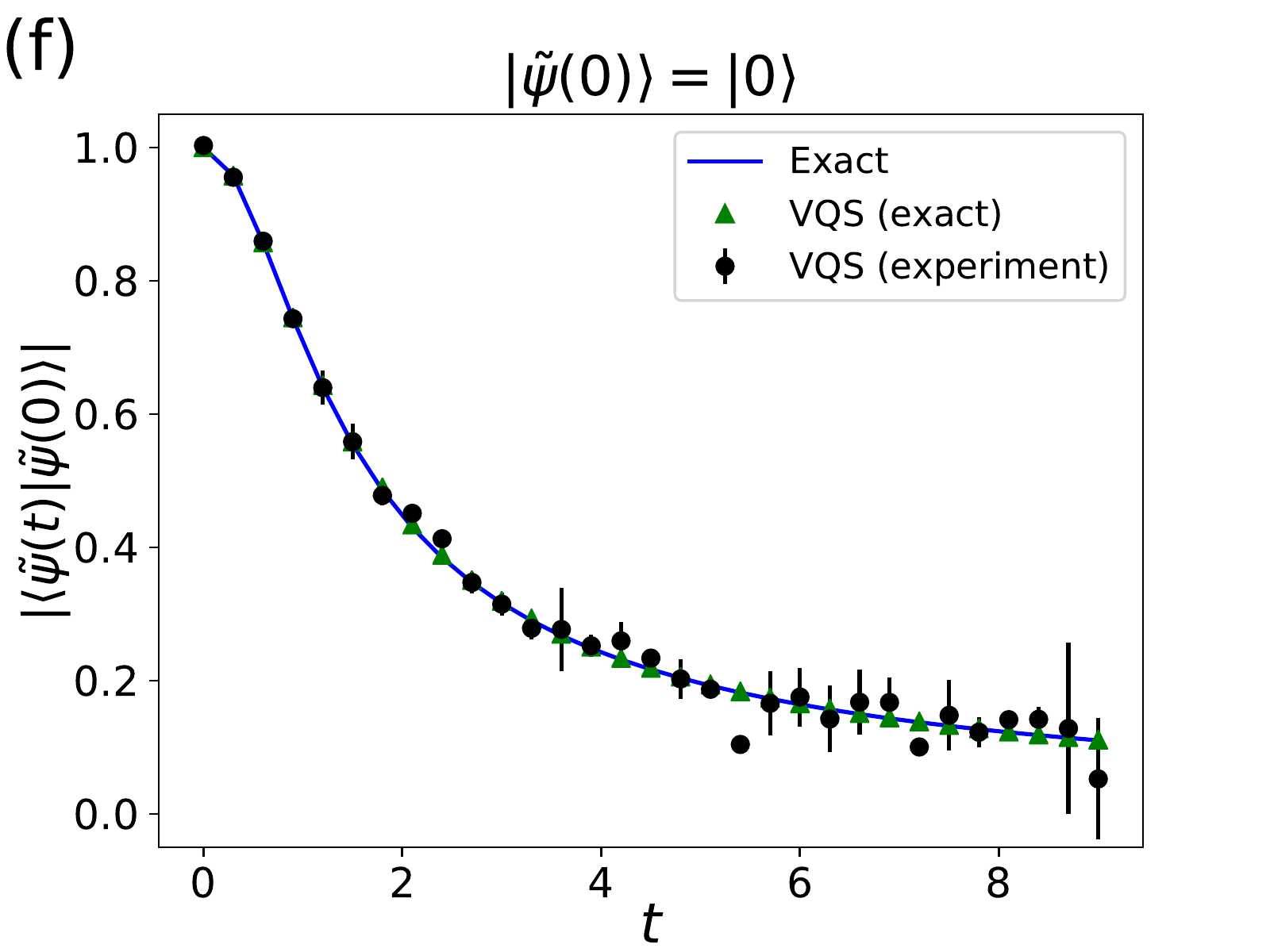}
    \caption{Result of running VQS on real quantum hardware.
    (a,c,e) Dynamics of the probability $P(t)$ in Eq.~\eqref{success-rate} calculated by the experimentally-obtained norm of the state, $\sqrt{\ip{\psi(t)}}$ (plotted in the insets).
    (b,d,f) Dynamics of the overlap the normalized state and the initial state, $|\ip{\tilde{\psi}(t)}{\psi_0}|$.
    ``Exact" indicates the exact solution of the Schr{\"o}dinger equation [Eq.~\eqref{eq: original schroedinger}] obtained by exact diagonalization, and ``VQS (exact)" does the dynamics obtained by solving the VQS equations [Eq.~\eqref{eq: MV}] in a numerically exact way.
    When there is no error in the experiment, ``VQS (exact)" matches with the experimental value ``VQS (experiment)".
    The difference between ``Exact" and "VQS(exact)" stems from the discretization error of the time that will vanish for $\delta t \to 0$. 
    All error bars in the panels are estimated from three independent runs of VQS experiments.
    }
    \label{fig: VQS result}
\end{figure*}

The results of the experiments are shown in Fig.~\ref{fig: VQS result}.
We consider the time step $\delta t= 0.15$ and  the second-order explicit Runge-Kutta method is used to calculate $\alpha(t+\delta t)$ and $\theta(t+\delta t)$ from $\alpha(t)$ and $\theta(t)$.
We plot the probability [Eq.~\eqref{success-rate}] by using the exact value of $\omega(H)=1/2$ and the experimentally obtained value of the norm parameter $\alpha = \sqrt{\ip{\psi(t)}}$ in the left column of Fig.~\ref{fig: VQS result}.
We also show the overlap $L(t)$ in the right column.
As evident from the figures, our experimental results agree well with the exact values and we again see the pseudospectral behavior in the case of the initial state $\ket{\tilde{\psi}(0)} = \frac{1}{\sqrt{2}}(\ket{0}-i\ket{1})$.
This illustrate the possibility to perform the simulation of the non-normal Hamiltonian dynamics on quantum computer by using VQS.

\section{Summary and outlook \label{sect: conclusion}}
In this paper, we have studied the non-normal Hamiltonian dynamics realized as the quantum trajectory with no quantum jump of the Lindblad master equation in open quantum systems and pointed out that the dynamics can reveal the nature of unconventional pseudospectrum of the non-normal Hamiltonian.
We have noticed the mathematics of the non-normal matrices:
in the presence of non-normality, it is known that not only the spectrum but also the pseudospectrum plays an important role in the linear dynamics of an unnormalized vector, especially its norm.
The norm in the case of quantum state dynamics has been related to the success probability for observing the continuous quantum trajectory of the Lindblad master equation.
We have formulated the transient suppression of the decay rate of the success probability due to the pseudospectral behavior and derived a non-Hermitian/non-normal analog of the time-energy uncertainty relation.
We have also discussed two methods to experimentally realize the non-normal dynamics and observe our theoretical findings on actual quantum computers: one uses a technique to realize non-unitary operations on quantum circuits and the other leverages a quantum-classical hybrid algorithm (VQS).
Our demonstrations using cloud-based quantum computers provided by IBM Quantum have shown the frozen dynamics of the norm in transient time, which can be regarded as a non-normal analog of the quantum Zeno effect.

As future work, it is interesting to perform experiments in larger systems (qubits) with employing sophisticated error mitigation techniques~\cite{Temme2017,Endo2018} to reduce the effect of the noise in near-term quantum computers.
It is also intriguing to investigate and verify the non-normal analog of the time-energy uncertainty relation, e.g., by controlling the parameters of a system and its pseudospectrum and calculating the overlap between the normalized state and an initial state.

\begin{acknowledgements}
This work was supported by JST CREST Grant No.~JPMJCR19T2, Japan. N.O. was supported by KAKENHI Grant No.~JP20K14373 from the JSPS.
The authors acknowledge the use of IBM Quantum services for this work. The views expressed are those of the authors, and do not reflect the official policy or position of IBM or the IBM Quantum team.
\end{acknowledgements}


\bibliography{NH-topo,YON}

\end{document}